\documentclass[useAMS,usenatbib]{mn2e}

\title[Stellar populations in a complete sample of local radio galaxies]{Stellar populations
in a complete sample of local radio galaxies}
\author[D. Raimann et al.]
  {D.~Raimann$^{1,2}$, T.~Storchi-Bergmann$^1$, H.~Quintana$^3$, R.~Hunstead$^4$
  \newauthor and L.~Wisotzki$^5$
\thanks{E-mail: raimann@if.ufrgs.br; thaisa@if.ufrgs.br; hquintana@astro.puc.cl;
rwh@physics.usyd.edu.au; lwisotzki@aip.de} \\
  $^1$ Instituto de F\'{\i}sica, Universidade Federal do Rio Grande do Sul, CP15051, Porto Alegre.
  91501-970, RS, Brazil\\
  $^2$ Universidade do Estado de Santa Catarina -- CEO, Rua Arcaj\'u s/n, Pinhalzinho,
  89870-000, SC, Brazil \\
  $^3$ Facultad de F\'{\i}sica, Pontificia Universitad Catolica de Chile, Santiago, Chile \\
  $^4$ School of Physics, University of Sydney, NSW 2006, Australia \\
  $^5$ Astrophysikalisches Institut Potsdam, Potsdam, Germany}

\begin{document}

\date{Accepted
      Received
      in original form }

\pagerange{\pageref{firstpage}--\pageref{lastpage}} \pubyear{}

\maketitle

\label{firstpage}

\begin{abstract}

We investigate the nature of the continuum emission and stellar
populations in the inner 1--3 kiloparsecs of a complete sample of
twenty-four southern radio galaxies, and compare the results with a
control sample of eighteen non-active early-type galaxies. Twelve of the
radio galaxies are classified as Fanaroff-Riley type I (FR\,I), eight as
FR\,II and four as intermediate or undefined type (FR\,x).  Optical
long-slit spectra are used to perform spectral synthesis as a function of
distance from the nucleus at an average sampling of 0.5--1.0\,kpc and
quantify the relative contributions of a blue featureless continuum and
stellar population components of different ages. Our main finding is a
systematic difference between the stellar populations of the radio and
control sample galaxies: the former have a larger contribution from an
intermediate age (1\,Gyr) component, suggesting a connection between the
present radio activity and a starburst which occurred $\sim$1\,Gyr
ago. In addition, we find a correlation between the contribution of the
1\,Gyr component and the radio power, suggesting that more massive
starbursts have led to more powerful radio emission. A similar relation is
found between the radio power and the mean age of the stellar population,
in the sense that stronger nuclear activity is found in younger galaxies.


We also find that the stellar populations of FR\,I galaxies are, on
average, older and more homogeneous than those of FR\,IIs. Significant
population gradients were found in only four radio galaxies, which are
also those with more than 10\,\% of their total flux at
4020\,\AA\ contributed by age components younger than 100\,Myr and/or a
featureless continuum (indistinguishable from a 3\,Myr old stellar
population).


\end{abstract}


\section{Introduction}

The nature of the UV-optical continuum in radio galaxies has been the
subject of a number of recent studies (Tadhunter et al.\ 1996, 2002;
Aretxaga et al.\ 2001; Wills et al.\ 2002, 2004). While at low redshifts
most radio galaxies seem to show stellar populations dominated in the
UV-optical by old stars, at high redshifts there is an UV excess,
frequently associated with structures that are aligned with the
large-scale radio structures, the so-called ``alignment effect'' (McCarthy
et al.\ 1987; Tadhunter et al.\ 1996, and references therein).

Several hypotheses have been advanced to explain the alignment effect:
episodes of recent star formation associated with the evolution of the
host galaxies (Lilly \& Longair 1984); star formation episodes triggered
by the passage of radio jets through the interstellar medium (Rees 1989);
and scattered light from a hidden quasar (Tadhunter et al.\ 1989; Fabian
1989). 

Using detailed spectropolarimetric observations, Tadhunter et al.\ (1996)
demonstrated that the UV-optical continuum of the low-redshift radio
galaxy 3C\,321 has a multi-component nature. At 3639\,\AA, an old stellar
population (15\,Gyr) contributes 34\,\% of the total flux, an intermediate
age population (1\,Gyr) contributes another 34\,\%, a hidden quasar
provides 22\,\% and a nebular continuum 10\,\%.  More recently, Tadhunter
et al.\ (2002) performed a similar study using a larger sample of 22
luminous radio galaxies at intermediate redshifts ($0.15 < z < 0.7$),
mostly composed of Fanaroff \& Riley (1974) class~II (FR\,II) radio
galaxies. All of them show a UV excess. These results emphasize the
multi-component nature of the UV continuum in radio galaxies: only
$\sim$1/3 comes from a significant contribution of polarized light
(scattered light from a hidden quasar), and the polarization level is
never larger than 10\,\%; at 3600\,\AA, the nebular continuum is present
in all objects, with varying proportions of 3--40\,\%;  direct AGN light
makes a significant contribution in 40\,\% of the objects and a
young/intermediate age stellar population (from 0.1 to 2\,Gyr)  is
significant in 15--50\,\% of the radio galaxies.


At lower redshifts, Aretxaga et al.\ (2001) and Wills et al.\ (2002, 2004)
have found similar results. Aretxaga et al.\ studied the optical spectra
of the nuclei of seven luminous nearby radio galaxies ($z<0.08$), which
mostly correspond to the FR\,II class. Three of them show a UV excess. One
is a broad-line radio galaxy where the UV excess is mainly due to
direct AGN light. In two cases, the blue spectrum is dominated by blue
supergiant and/or giant stars with ages from 7 to 40\,Myr. Wills et al.\
(2002) studied the optical spectra of nine FR\,II radio galaxies ($0.05 <
z < 0.2$). Four galaxies display a UV excess, with one being a broad-line
radio galaxy.  In the other three, the UV excess is due to young and/or
intermediate age stellar populations (from 0.5 to 2\,Gyr).  The four radio
galaxies without UV excess have stellar populations typical of elliptical
galaxies. The contribution of the nebular continuum varies from 0 to
26\,\% of the total flux at 3660\,\AA. No significant contribution from
polarized light was found.  In a subsequent paper, Wills et al.\ (2004)
performed a similar study on 12 low luminosity FR\,I radio galaxies
($z <0.2$), finding that three objects show UV excess, the main
contribution being due to young and/or intermediate age stars.

The studies so far available in the literature
refer to samples dominated by FR\,II galaxies,
except for that of Wills et al. (2004) which is constrained
to FR\,I galaxies only.  There are no previous studies
including a control sample, only a few stellar population
studies of early-type galaxies (e.g. Quintana et al. 1990).


The novelty of our present work is three-fold:

(1) To minimize selection effects we chose a sample limited in redshift
and radio flux, which is complete in the sense that it comprises the
closest most luminous radio galaxies. It contains both FR\,I and FR\,II
radio galaxies.

(2) We defined a control sample of early-type galaxies in order to look
for systematic differences between the radio galaxies and non-active
galaxies of similar Hubble types.

(3) We extended the stellar population studies out to a few kiloparsecs
from the nucleus, at a sampling of $\sim$0.2--1 kpc.


Our goal was to apply the technique we successfully used in previous
studies of stellar populations of Seyfert, LINER and non-active
galaxies (Cid-Fernandes et al.\ 1998; Raimann et al.\ 2001, 2003, 2004)
to address the following questions: 
\begin{itemize}
\item What fraction of radio galaxies
show a UV excess when compared with non-active galaxies of
the same Hubble type? \item What is the nature of this UV excess?
\item What fraction of radio galaxies exhibit signatures of recent star 
formation?
\item Are there systematic differences between the stellar population
of FR\,I and FR\,II radio galaxies? \item Are there systematic differences
between the stellar population of radio galaxies and non-active galaxies of the
same Hubble type?
\end{itemize}


The paper is organized as follows: In Section 2 we describe the sample
galaxies and the observations. In Section 3 we present the measurements of
continuum colours, line equivalent widths and their radial variations. We
describe the method and results of spectral synthesis in Section 4. In
Section 5 we discuss the results, and in Section 6 we present our
conclusions.

\section{Sample and Observations}

\subsection{Radio galaxies}

The sample of radio galaxies comprises 24 objects with $z<0.08$ and
integrated radio flux density $S(408\,\mathrm{MHz})>4.0$\,Jy, extracted
from the Molonglo Southern 4\,Jy sample (Burgess \& Hunstead, 1994, 2005),
with declinations in the range $-85^{\circ}< \delta <-30^{\circ}$ and
galactic latitudes $|b| > 10^{\circ}$. The complete sample according to
these criteria consists of 30 objects, but we could not observe
6 sources due to poor observational
conditions. However, the exclusion of these sources does not
seem to bias the survey or affect our conclusions.

In Table \ref{amostrarg} we list the position, morphological type, radio
classification (Fanaroff \& Riley 1974), emission-line class (see below),
apparent magnitude $B$, $S(408\,\mathrm{MHz})$, radial velocity $cz$ and
foreground galactic reddening $E(B-V)_\mathrm{Gal}$, for each galaxy. 12
radio galaxies are classified as Fanaroff-Riley I (FR\,I), eight are
FR\,II and four are of intermediate or undefined type (FR\,x). The
emission-line classes have been assigned according to the following
criteria: BLRG are radio galaxies with broad emission lines (i.e.
permitted emission lines have both narrow and broad-line
components);  NLRG have narrow emission lines
with equivalent widths $W_\lambda > 5$\,\AA; WLRG
are the ones with weak emission lines ($W_\lambda<5$\,\AA); and NO-E have
no emission lines. The radial velocities $cz$ range from 8400 to
22500\,km\,s$^{-1}$, with a mean value of 15000\,km\,s$^{-1}$ ($z\simeq
0.05$).  The data were extracted from the NASA/IPAC Extragalactic Database
(NED)\footnote{The NASA/IPAC Extragalactic Database (NED) is operated by
the Jet Propulsion Laboratory, California Institute of Technology, under
contract with the National Aeronautics and Space Administration.}.

Long-slit spectra of these galaxies were obtained with the EMMI
spectrograph at the 3.5m New Technology Telescope (NTT) of the European
Southern Observatory (ESO) at La Silla in 2001 September and 2002
February. The spectra were obtained in two segments, covering the
wavelength ranges 3500--5000\,\AA\ and 4800--7300\,\AA, at spectral
resolutions of 5 and 3.6\,\AA, respectively. Two exposures were obtained
at each spectral region in order to eliminate cosmic rays, yielding total
exposure times of 1800\,s in the blue region and 3600\,s in the red.  The
slit, with a width corresponding to 1.5$^{\prime\prime}$ on the sky, was
oriented along the major axis of the extended radio emission. The mean
spatial scale is $\sim 1\,\mathrm{kpc}/\mathrm{arcsec}$. We present a log
of the observations in Table \ref{amostrarg2}.


The two-dimensional spectra were combined and reduced using standard tasks
in IRAF (involving bias and flat-field correction, wavelength and flux
calibration).  One-dimensional spectra were extracted in windows of 1.5
arcsec in the bright nuclear regions and progressively larger windows
towards the fainter outer regions. The red spectra were smoothed
lightly to match the lower resolution of the blue spectra, after
which the two segments were combined. Finally, the resulting spectra were
corrected for foreground galactic extinction and shifted to the rest
frame.

\begin{figure*}
\vspace{16cm}
\caption{Sample of nuclear and extranuclear spectra of the radio galaxies:
(a) a BLRG; (b) a NLRG; (c) a WLRG and (d) a radio galaxy without emission lines.}
\label{spectraradio}
\includegraphics{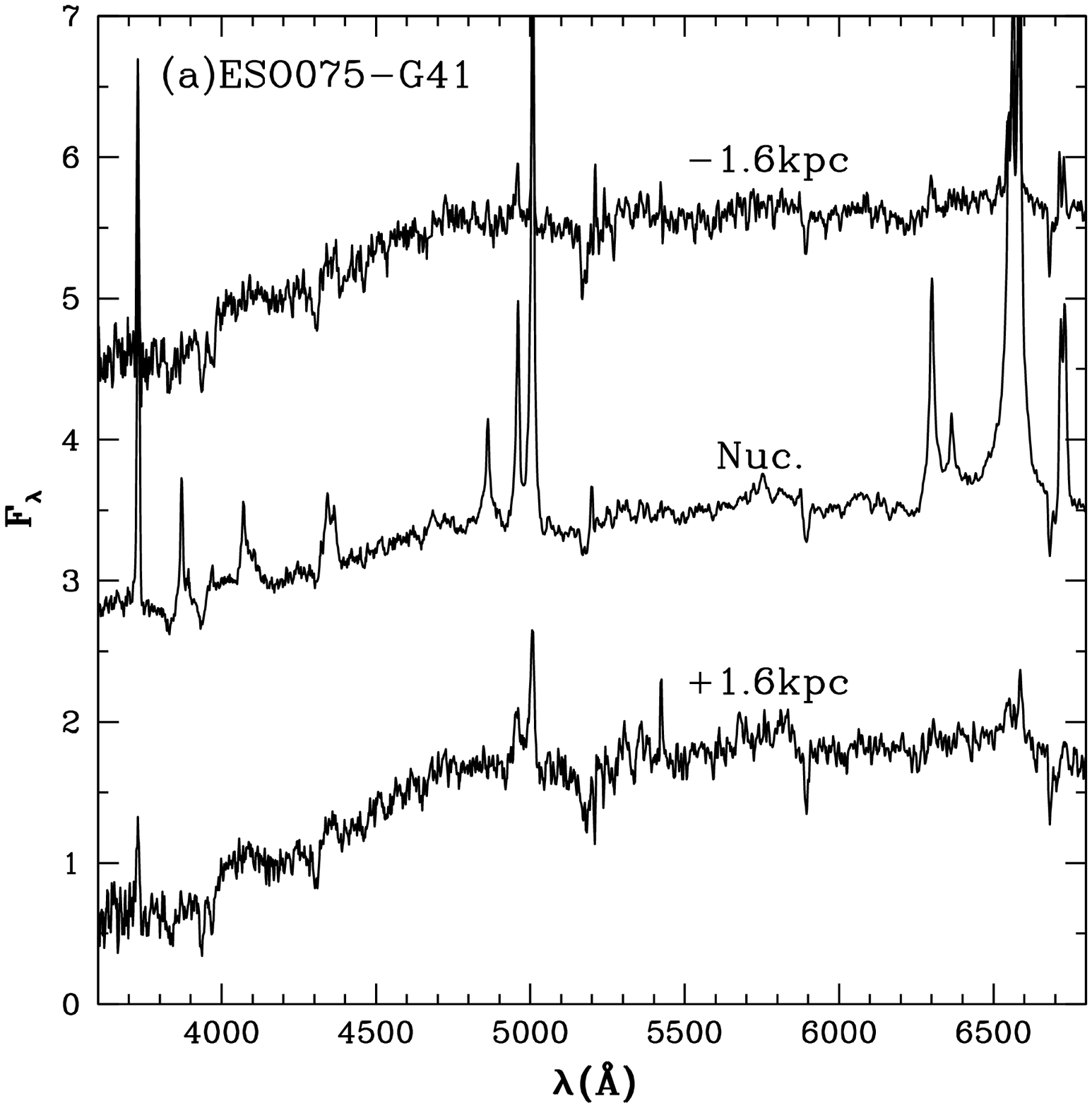}
\includegraphics{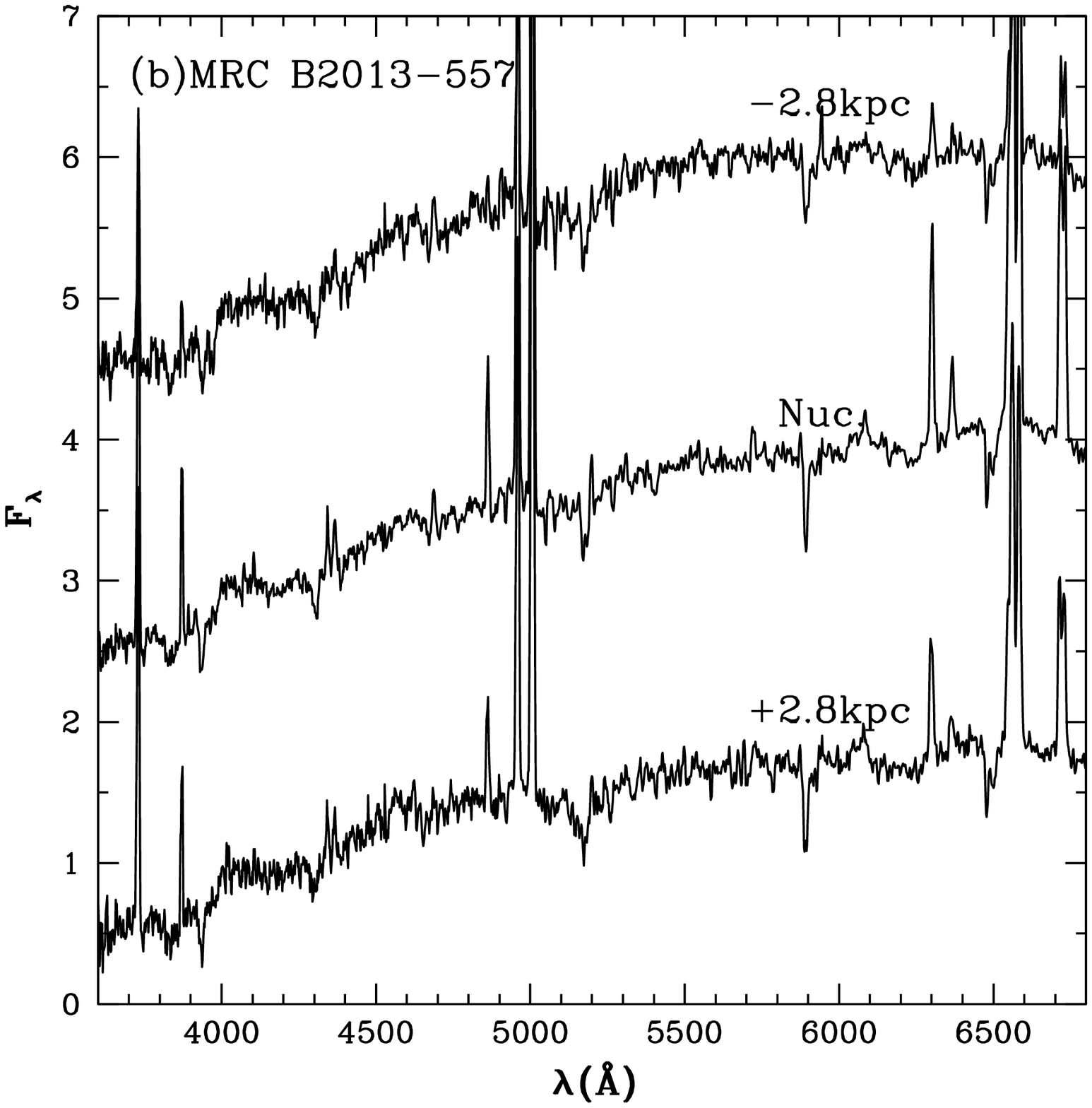}
\includegraphics{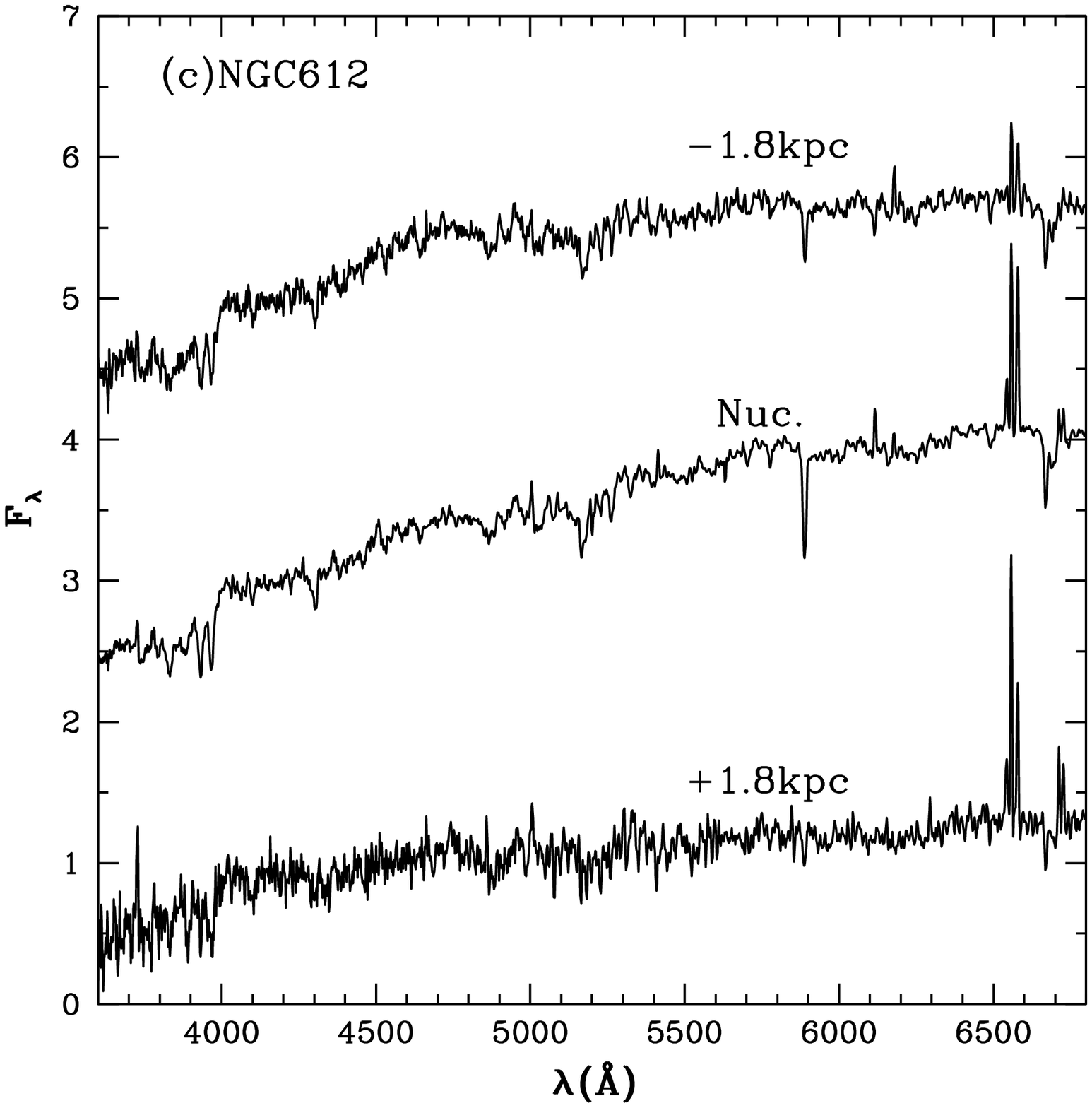}
\includegraphics{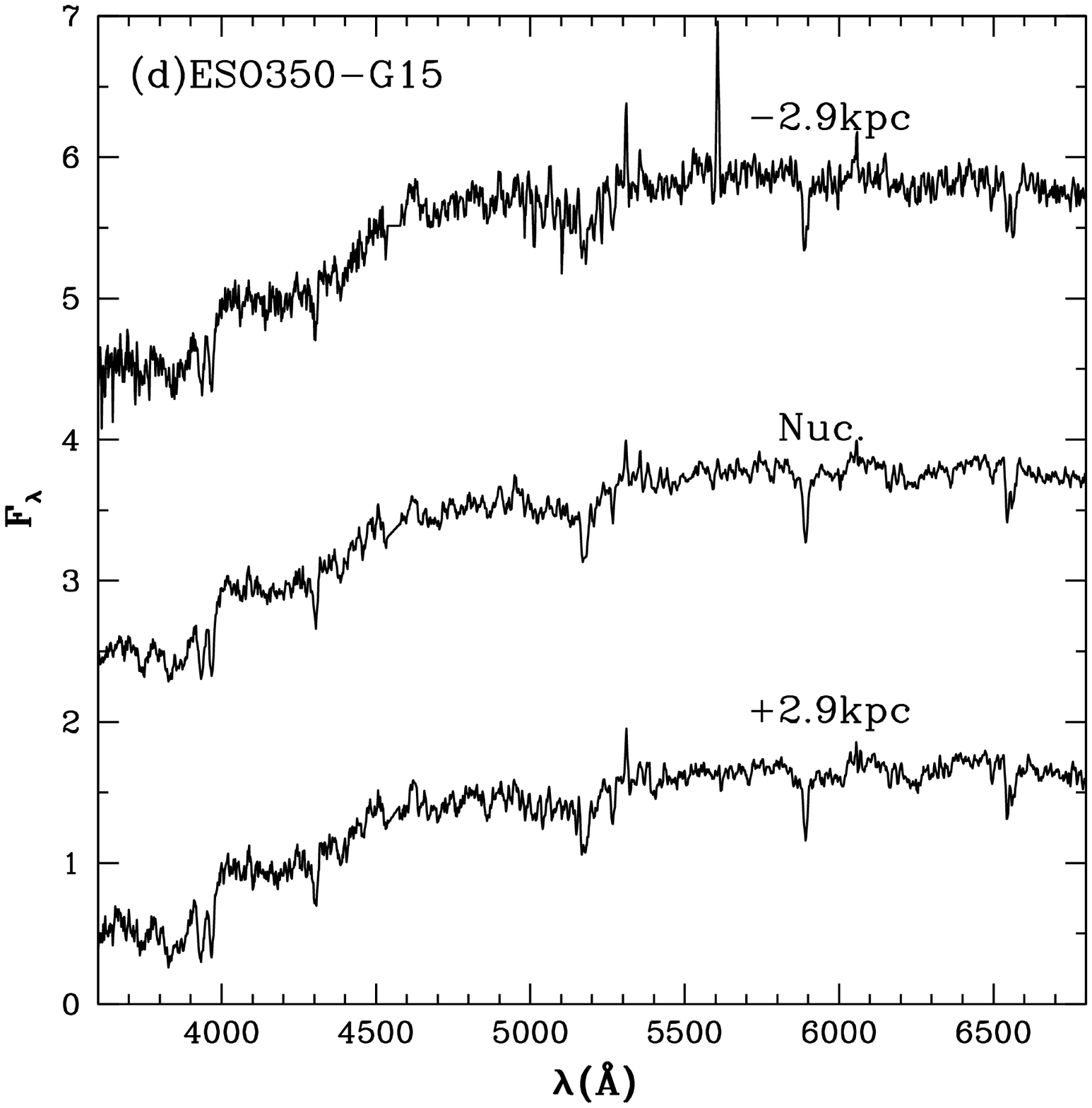}
\end{figure*}

The spatial coverage ranges between 1 and 7\,kpc at the galaxies
(1.5$^{\prime\prime}$--8$^{\prime\prime}$). The signal-to-noise (S/N)
ratios of the extracted spectra range between 10 and 30. Representative
spectra of the sample are shown in Fig.\ \ref{spectraradio}.

\begin{table*}
\caption{The radio galaxy sample.}
\label{amostrarg}
\begin{center}
\begin{tabular}{llcccccrrrc} \hline
Name & ID & $\alpha$(2000) & $\delta$ (2000) & Morph.$^1$
& FR$^2$ & Emission & \multicolumn{1}{c}{$B$} &
\multicolumn{1}{c}{$S_{408}$} & \multicolumn{1}{c}{$cz$} &
\multicolumn{1}{c}{$E(B-V)_\mathrm{Gal}$} \\
 & & & & type & type & type & (mag) & (Jy) & (km/s) & (mag) \\ \hline
MRC B0023$-$333 & ESO\,350-G15    & 00 25 31 & $-$33 02 46 & E3          & I    & NO-E & 14.6    & 4.1 & 14940    & 0.015 \\
MRC B0131$-$367 & NGC\,612         & 01 33 57 & $-$36 29 35 & SA0 pec     & II   & WLRG & 13.5    & 17.1 & 8925    & 0.020 \\
MRC B0214$-$480 & ESO\,198-G1     & 02 16 46 & $-$47 49 23 & E4          & I    & WLRG & 15.0    & 5.5 & 19190    & 0.023 \\
MRC B0319$-$454 & ESO\,248-G10    & 03 21 08 & $-$45 12 51 & S           & II   & NLRG & 15.6    & 8.3 & 18887    & 0.015 \\
MRC B0332$-$391 & --             & 03 34 07 & $-$39 00 03 & E           & I    & NO-E & 15.7    & 4.2 & 18680    & 0.019 \\
MRC B0344$-$345 & --             & 03 46 30 & $-$34 22 45 & E           & I/II & NLRG & 16.2    & 7.6 & 16130    & 0.012 \\
MRC B0427$-$539 & IC\,2082         & 04 29 07 & $-$53 49 39 & --          & I    & WLRG & 14.0    & 14.6 & 12351    & 0.012 \\
MRC B0429$-$616 & --             & 04 30 22 & $-$61 32 01 & --          & I    & WLRG & 15.5    & 4.4 & 16680    & 0.022 \\
MRC B0456$-$301 & --             & 04 58 26 & $-$30 07 22 & E3          & x    & NLRG & 17.5    & 7.2 & 18900    & 0.013 \\
MRC B0518$-$458 & Pictor\,A       & 05 19 49 & $-$45 46 44 & (R')SA0 pec & II   & BLRG & 16.6    & 166.0 & 10510    & 0.043 \\
MRC B0618$-$371 & ESO\,365-IG6    & 06 20 00 & $-$37 11 42 & SAB0-:      & II   & NO-E & 14.8    & 5.8 & 9838    & 0.080 \\
MRC B0620$-$526 & --             & 06 21 43 & $-$52 41 36 & --          & I    & WLRG & 15.5    & 9.3 & 15320    & 0.068 \\
MRC B0625$-$536$^3$ & ESO\,161-IG7   & 06 26 20 & $-$53 41 33 & E pec       & I    & NO-E & 14.9    & 26.0 & 16507    & 0.094 \\
MRC B0715$-$362 & --             & 07 17 08 & $-$36 22 00 & SA0-        & I    & NO-E & 15.7    & 5.7 & 9593    & 0.282 \\
MRC B1123$-$351 & ESO\,377-G46    & 11 25 52 & $-$35 23 41 & (R)SAB(rs)0 & I    & NO-E & 14.0    & 6.6 & 10119    & 0.087 \\
MRC B1407$-$425 & ESO\,271-G20    & 14 10 28 & $-$42 46 56 & S           & x    & WLRG & 15.1    & 4.7 & 15889    & 0.081 \\
MRC B1413$-$364 & --             & 14 16 33 & $-$36 40 54 & E           & II   & NLRG & 17.7    & 5.7 & 22394    & 0.066 \\
MRC B1637$-$771 & --             & 16 44 16 & $-$77 15 48 & --          & II   & NLRG & 16.3    & 13.5 & 12801    & 0.099 \\
MRC B1929$-$397 & ESO\,338-IG11   & 19 33 23 & $-$39 40 23 & --          & II?  & WLRG & 15.5    & 4.3 & 22504    & 0.152 \\
MRC B2013$-$557 & --             & 20 18 01 & $-$55 39 30 & E           & I    & NLRG & 16.2    & 4.8 & 18000    & 0.066 \\
MRC B2148$-$555 & --             & 21 51 29 & $-$55 20 13 & E2          & I    & NO-E & 14.8    & 5.8 & 11627    & 0.024 \\
MRC B2152$-$699 & ESO\,075-G41    & 21 57 06 & $-$69 41 23 & SA0-        & II   & BLRG & 14.1    & 61.6 & 8476    & 0.029 \\
MRC B2158$-$380 & AM\,2158$-$380     & 22 01 17 & $-$37 46 25 & Sa          & II   & NLRG & 14.8    & 4.1 & 9983    & 0.018 \\
MRC B2354$-$350 & ESO\,349-G10    & 23 57 00 & $-$34 45 30 & E4          & I    & NLRG & 14.0    & 8.7 & 14705    & 0.013 \\
\hline
\end{tabular}
\end{center}
\begin{flushleft}
$^1$ Galaxy  morphology from NED\\
$^2$ Radio type according to Fanaroff \&  Riley (1974)\\
$^3$ Dumbell galaxy; radio-source associated with eastern member\\
\end{flushleft}
\end{table*}

\begin{table*}
\caption{Observing log for the radio galaxies and spatial scale.}
\label{amostrarg2}
\begin{center}
\begin{tabular}{lrrcccrc} \hline
Name           & P.A.($^{\circ}$)$^{1}$ & $\psi(^{\circ})^{2}$ &
Air mass & Scale $^{3}$ &
Seeing ($^{\prime\prime}$) & Linear$^{3,4}$ & Log L$_{408}$ \\
&   &    &   &(kpc/$^{\prime\prime}$) &  &(kpc) & \\\hline
ESO\,350-G15    & 11 & 106 & 1.02 & 0.91 &  1.2 & 62 & 25.29 \\
NGC\,612         & 103 & 105 & 1.07 & 0.56 & 1.2 & 304 & 25.47 \\
ESO\,198-G1     & 173 & 148 & 1.07 & 1.15 & 1.2 & 490 & 25.63\\
ESO\,248-G10    & 49 & 5 & 1.04 & 1.13 & 1.1 & 1745 & 25.80 \\
MRC B0332$-$391    & 100 & 80 & 1.15 & 1.12 & 1.1 & 470 & 25.50 \\
MRC B0344-345      & 105 & 92 & 1.13 & 0.98 & 1.5 & 265 & 25.62 \\
IC\,2082         & 109 & 71 & 1.26 & 0.76 & 1.2 & 210 & 25.64 \\
MRC B0429$-$616    & 14 & 36 & 1.25 & 1.01 & 1.2 & 115 & 25.41 \\
MRC B0456$-$301    & 103 & 83 & 1.14 & 1.13 & 1.0 & 395 & 25.73 \\
Pictor\,A       & 102 & 82 & 1.35 & 0.65 & 1.0 & 283 & 26.57 \\
ESO\,365-IG6    & 88 & 110 & 1.05 & 0.61 & 1.0 & 60 & 25.07 \\
MRC B0620$-$526    & 50 & 66 & 1.25 & 0.93 & 1.2 & 296 & 25.65 \\
ESO\,161-IG7    & 103 & 65 & 1.29 & 1.00 & 0.9 & 119 & 26.16 \\
MRC B0715$-$362    & 82 & 90 & 1.22 & 0.60 & 0.7 & 269 & 25.04 \\
ESO\,377-G46    & 87 & 71 & 1.05 & 0.63 & 0.9 & 38 & 25.11 \\
ESO\,271-G20    & 102 & 117 & 1.18 & 0.96 & 1.2 & 47 & 25.40 \\
MRC B1413$-$364    & 29 & 109 & 1.05 & 1.32 & 1.0 & 252 & 25.78 \\
MRC B1637$-$771    & 0 & 40 & 1.60 & 0.79 & 1.4 & 206 & 25.66 \\
ESO\,338-IG11   & 134 & 157 & 1.02 & 1.33 & 0.7 & 150 & 25.66 \\
MRC B2013$-$557    & 155 & 150 & 1.13 & 1.08 & 1.3 & 1300 & 25.52 \\
MRC B2148$-$555    & 26 & 164 & 1.12 & 0.72 & 1.2 & 560 & 25.20 \\
MRC ESO\,075-G41    & 18 & 5 & 1.31 & 0.53 & 0.7 & 42 & 25.97 \\
MRC AM\,2158$-$380   & 40 & 140 & 1.03 & 0.62 & 1.2 & 71 & 24.93 \\
MRC ESO\,349-G10    & 141 & 123 & 1.01 & 0.90 & 1.4 & 61 & 25.59 \\
\hline
\end{tabular}
\end{center}
\begin{flushleft}
$^{1}$ Slit position angle\\
$^{2}$ Parallactic angle\\
$^{3}$ Calculated for a flat WMAP cosmology with $H_0=75$
km\,s$^{-1}$\,Mpc$^{-1}$\\
$^{4}$ Linear extent of the radio source\\
\end{flushleft}
\end{table*}

\begin{table*}
\caption{The control sample.}
\label{amostranon1}
\begin{center}
\begin{tabular}{lccccrc} \hline
Name           & $\alpha$(2000) & $\delta$ (2000)     & Morph. & $B$     & $cz$     & $E(B-V)_\mathrm{Gal}$ \\
               &                &                     & type   & (mag) & (km/s) & (mag) \\ \hline
NGC1404        & 03 38 52 & $-$35 35 37 & E1            & 10.35 & 1497 &   0.011 \\
NGC1700        & 04 56 56 & $-$04 51 55 & E4            & 12.20 & 3895 &   0.043 \\
NGC2865        & 09 23 30 & $-$23 09 43 & E3-4          & 12.57 & 2627 &   0.082 \\
NGC3091        & 10 00 14 & $-$19 38 13 & E3:           & 12.13 & 3964 &   0.043 \\
NGC3585        & 11 13 17 & $-$26 45 18 & E7/S0         & 10.88 & 1399 &   0.064 \\
NGC3706        & 11 29 44 & $-$36 23 33 & SA(rs)0-      & 12.38 & 2977 &   0.092 \\
NGC3904        & 11 49 12 & $-$29 16 35 & E2-3:         & 11.83 & 1496 &   0.072 \\
NGC3923        & 11 51 01 & $-$28 48 22 & E4-5:         & 10.80 & 1788 &   0.083 \\
NGC4373        & 12 25 18 & $-$39 45 37 & SAB(rs)0-:    & 11.90 & 3396 &   0.080 \\
NGC4825        & 12 57 12 & $-$13 39 53 & SA(0)         & 12.63 & 4452 &   0.049 \\
NGC4936        & 13 04 17 & $-$30 31 31 & E0            & 11.77 & 3117 &   0.083 \\
NGC5061        & 13 18 04 & $-$26 50 11 & E0            & 11.44 & 2661 &   0.068 \\
NGC5328        & 13 52 53 & $-$28 29 16 & E1:           & 12.67 & 4740 &   0.062 \\
NGC5813        & 15 01 11 & $-$01 42 08 & E1-2          & 11.45 & 1972 &   0.057 \\
NGC6684        & 18 48 57 & $-$65 10 26 & (L)SB(r)0$^+$ & 11.31 & 847 &    0.067 \\
NGC6861        & 20 07 19 & $-$48 22 12 & SA(s)0        & 12.12 & 2819 &   0.054 \\
NGC7049        & 21 19 00 & $-$48 33 43 & SA(s)0        & 11.72 & 2231 &   0.007 \\
NGC7079        & 21 32 35 & $-$44 04 00 & (L)SB(r)0     & 12.46 & 2670 &   0.031 \\ \hline
\end{tabular}
\end{center}
\end{table*}

\begin{table*}
\caption{Log of observations of the control sample.}
\label{amostranon2}
\begin{tabular}{lccccccc} \hline
Name           & Exp. time(s) & P.A.($^{\circ}$) & $\psi(^{\circ})$ & Air mass & kpc/$^{\prime\prime}$
& Telescope & Seeing ($^{\prime\prime}$) \\ \hline
NGC1404        & 5400 & 90 & 100 & 1.50 & 0.10 & 1.5m ESO & 0.8 \\
NGC1700        & 600  & 20 & 13 & 1.10 & 0.25 & 3.6m NTT  & 1.5 \\
NGC2865        & 5400 & 90 & 36 & 1.02 & 0.17 & 1.5m ESO  & 1.0 \\
NGC3091        & 5400 & 90 & 65 & 1.15 & 0.25 & 1.5m ESO  & 0.9 \\
NGC3585        & 5400 & 90 & 150 & 1.01 & 0.09 & 1.5m ESO & 1.0 \\
NGC3706        & 5400 & 90 & 90 & 1.20 & 0.19 & 1.5m ESO  & 0.9 \\
NGC3904        & 5400 & 90 & 81 & 1.19 & 0.10 & 1.5m ESO  & 0.8 \\
NGC3923        & 5400 & 90 & 96 & 1.00 & 0.12 & 1.5m ESO  & 0.8 \\
NGC4373        & 5400 & 90 & 20 & 1.02 & 0.21 & 1.5m ESO  & 0.8 \\
NGC4825        & 5400 & 90 & 30 & 1.05 & 0.28 & 1.5m ESO  & 0.9 \\
NGC4936        & 5400 & 90 & 90 & 1.01 & 0.19 & 1.5m ESO  & 1.0 \\
NGC4936        & 600  & 86 & 86 & 1,07 & 0.19 & 3.6m NTT  & 1.0 \\
NGC5061        & 5400 & 90 & 78 & 1.10 & 0.17 & 1.5m ESO  & 0.9 \\
NGC5328        & 5400 & 90 & 95 & 1.10 & 0.30 & 1.5m ESO  & 0.8 \\
NGC5813        & 600  & 90 & 30 & 1,25 & 0.13 & 3.6m NTT  & 1.0 \\
NGC6684        & 1800 & 155 & 155  & 1.26 & 0.05 & 4m CTIO & -- \\
NGC6861        & 1800 & 125 & 125  & 1.14 & 0.18 & 4m CTIO & -- \\
NGC7049        & 1800 & 116 & 116  & 1.19 & 0.14 & 4m CTIO & -- \\
NGC7079        & 5400 & 90 & 50  & 1.10 & 0.17 & 1.5m ESO & 1.5 \\ \hline
\end{tabular}
\end{table*}

\begin{figure*}
\vspace{8cm}
\caption{Spectra of an elliptical (a) and a lenticular (b) galaxy from the
control sample.}
\label{spectranon}
\includegraphics{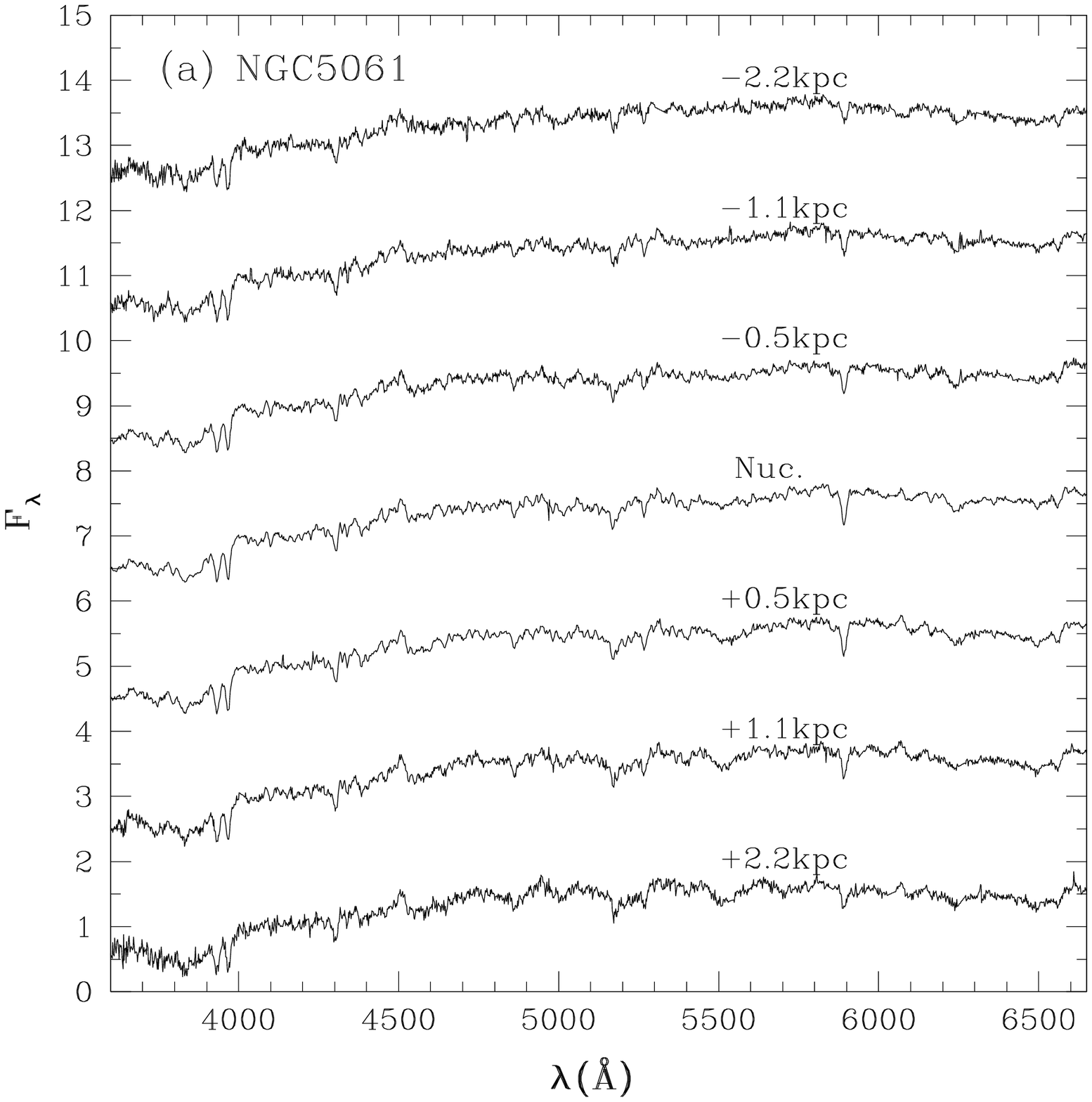}
\includegraphics{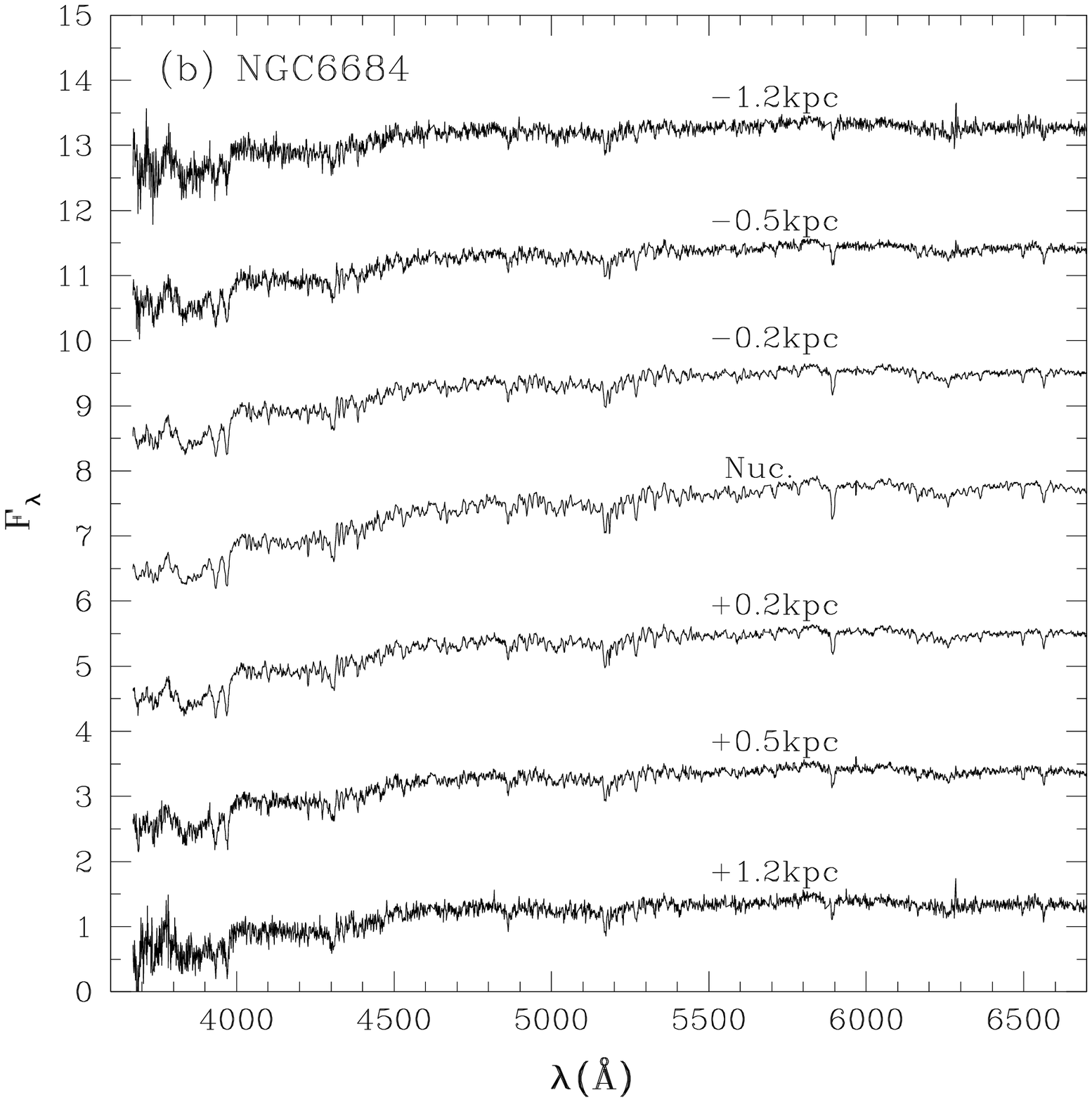}
\end{figure*}

\subsection{Control sample}

The control sample comprises 18 non-active early-type galaxies:  7
lenticulars and 11 ellipticals. They were selected to have
similar Hubble types and absolute magnitudes to those of the radio galaxies,
and no signs of nuclear activity. In order to be
observable with a smaller telescope, the control sample galaxies were
selected to be closer ($z<0.015$) than the radio galaxies.
As we have not used any particular spectral characteristic 
to build this sample, we believe it is not biased,
i.e., the conclusions of this work do not depend on the
choice of these galaxies as control sample.

In Table \ref{amostranon1} we list the positions, morphological types,
apparent magnitudes $B$, radial velocities and foreground galactic
reddening values of the control sample.

Long-slit spectra of the control sample were obtained mostly with the
Boller \& Chivens spectrograph at the 1.52\,m telescope at ESO. A few
spectra were obtained with the the Cassegrain spectrograph at the 4\,m
Blanco telescope at Cerro Tololo Interamerican Observatory and with the
EMMI spectrograph at the 3.5\,m NTT at ESO. The wavelength range covered
was 3600--7000\AA, at a spectral resolution of 4--6\,\AA. The slit, with
a width corresponding to 1.5$^{\prime\prime}$ on the sky, was oriented
along the parallactic angle.

The galaxy NGC\,4936 was observed with both the 1.52m telescope and the
NTT in order to check for any systematic differences in the spectra due
to different instrumentation. We found none; the results of the
measurements and population synthesis are identical for the two datasets
within the errors. In order to illustrate this, we kept the entries
corresponding to both observations of this galaxy in all the tables of
the paper.

A log of these observations is presented in Table \ref{amostranon2},
where we list the exposure time, the slit position angle, the parallactic
angle, air mass, spatial scale, telescope used, and seeing. The mean
spatial scale in the control sample is 0.16\,kpc/arcsec, a factor 
$\sim$6 smaller than for the radio galaxy sample.

Two or three exposures of each galaxy were obtained in order to eliminate
cosmic rays. The two-dimensional spectra were combined and reduced using
standard tasks in IRAF, as for the radio galaxy sample. One-dimensional
spectra were extracted in windows of 1.5--3.2 arcsec in the bright
nuclear regions and progressively larger windows towards the fainter
outer regions.

The spatial coverage ranged between 0.15 and 4\,kpc (3--36 arcsec)
from the nucleus. The S/N ratio of the extracted spectra ranges between 10
and 30. Fig.\ \ref{spectranon} shows nuclear and extranuclear
spectra of two representative galaxies of the control sample.

Note that the extraction samples smaller regions at the non-active
galaxies than at the radio galaxies. Thus, whenever we compare the
results for the radio galaxies with those for the control sample
galaxies, we combine the nuclear and a few extra-nuclear extractions in
the latter in order to cover similar spatial extents in the two
samples ($\sim 1$\,kpc).

\section{Equivalent widths of absorption features and continuum colours}

The analysis of the stellar population properties was performed using the
same principles as in our previous papers (e.g. Raimann et al.\ 2001,
2003). We constructed a pseudo-continuum at selected pivot-points of the
spectra and measured the equivalent widths (W$_\lambda$) of eight
absorption features. The pivot-points for the continuum are at rest
wavelengths 3660, 3780, 4020, 4510, 4630, 5313, 5870, 6080, and
6630\,\AA\ and the absorption features we measured are as follows: WLB (a
blend of weak lines in the near-UV, within the spectral window
$\lambda\lambda$3810--3822\,\AA), H9 (a blend of absorption lines which
includes H9, window $\lambda\lambda$3822--3858\,\AA), Ca\,{\sc ii}~K
($\lambda\lambda$3908--3952\,\AA), Ca\,{\sc ii}~H+H$\epsilon$
($\lambda\lambda$3952--3988\,\AA), the CN band
($\lambda\lambda$4150--4214\,\AA), the G band
($\lambda\lambda$4284--4318\,\AA), Mg\,{\sc i}+MgH
($\lambda\lambda$5156--5196\,\AA) and Na\,{\sc i}
($\lambda\lambda$5880--5914\,\AA).

Equivalent widths and continuum definitions are based on Bica \& Alloin
(1986), Bica (1988), Bica, Alloin \& Schmitt (1994) and Raimann et al.\
(2001). The use of the same set of pivot-points and wavelength windows
allows a detailed quantitative analysis of the stellar populations via
synthesis techniques, using the spectral library of star clusters of Bica
\& Alloin (1986) and Bica (1988).

\subsection{Nuclear values}

A summary of the nuclear measurements for the radio galaxies is presented
in Table \ref{var}. This table lists the range of equivalent widths and
continuum fluxes (relative to that at $\lambda$4020) measured in the
nuclear spectra of the radio galaxies.  The table also shows, for
comparison, the corresponding values for the control sample. For the
latter galaxies, nuclear and extranuclear spectra have been combined to
cover a spatial extent $\sim$1\,kpc, similar to that covered by the
nuclear extractions of the radio galaxies. It can be seen that, while the
upper limits for the two sub-samples are similar, the radio galaxies have
smaller $W_\lambda$ on average, indicating the presence of an
excess blue continuum.  The radio galaxies must either have stellar
populations younger or more metal poor than those of the control sample,
or nuclear spectra that are diluted by an AGN continuum.

The range of continuum fluxes is also broader in the radio sample, for
which both bluer and redder continua are observed relative to the control
sample.

\begin{table*}
\caption{Range of nuclear equivalent widths and continuum fluxes from
each sub-sample.}
\label{var}
\begin{center}
\begin{tabular}{lcccccccc} \hline
Sample      & $W_\mathrm{WLB}$ & $W_\mathrm{H9}$ & $W_\mathrm{Ca\,{\small II}~K}$ &
  $W_\mathrm{Ca\,{\small II}~H+H\epsilon}$ & $W_\mathrm{CN}$ &
  $W_\mathrm{G}$ & $W_\mathrm{Mg\,{\small I}+MgH}$ & $W_\mathrm{Na\,{\small I}}$ \\ \hline
Radio galaxies          & 1--7 & 7--18 & 8--20 & 8--15 & 5--15  & 4--12  & 4--11 & 4--9 \\
Non-active galaxies     & 4--7 & 13--20 & 14--20 & 12--14 & 7--12  & 10--19 & 6--12 & 3--7 \\ \hline
                        & 3660\AA & 4510\AA & 5870\AA & 6630\AA & & & & \\ \hline
Radio galaxies          & 0.47--1.22 & 0.83--1.57 & 0.74--3.24 & 0.67--3.79 & & & & \\
Non-active galaxies     & 0.52--0.63 & 1.25--1.69 & 1.62--2.41 & 1.51--2.40 & & & & \\ \hline
\end{tabular}
\end{center}
\end{table*}

\begin{figure*}
\vspace{15cm}
\caption{Radial variations of equivalent width, continuum colour and
surface brightness for radio galaxies (top) and control sample galaxies
(bottom). The first panel, from top to bottom, shows $W_\mathrm{WLB}$
(solid line)  and $W_\mathrm{H9}$ (dotted), the second shows
$W_\mathrm{Ca\,{\small II}~K}$ (solid) and $W_\mathrm{Ca\,{\small
II}~H+H\epsilon}$ (dotted), the third, $W_\mathrm{G band}$ (solid) and
$W_\mathrm{CN band}$ (dotted), the fourth, $W_\mathrm{Mg\,{\small
I}+MgH}$ (solid) and $W_\mathrm{Na\,{\small I}}$ (dotted). The fifth
panel shows the continuum flux ratio between 5870 and 4020\,\AA.  The
sixth panel shows the run of the surface brightness at 4020\,\AA\ (in
units of $10^{-15}$ erg cm$^{-2}$ s$^{-1}$ \AA$^{-1}$ arcsec$^{-2}$)
along the slit. The dotted and dashed vertical lines mark distances of
1\,kpc and 3\,kpc from the nucleus, respectively.}
\label{variation}
\includegraphics{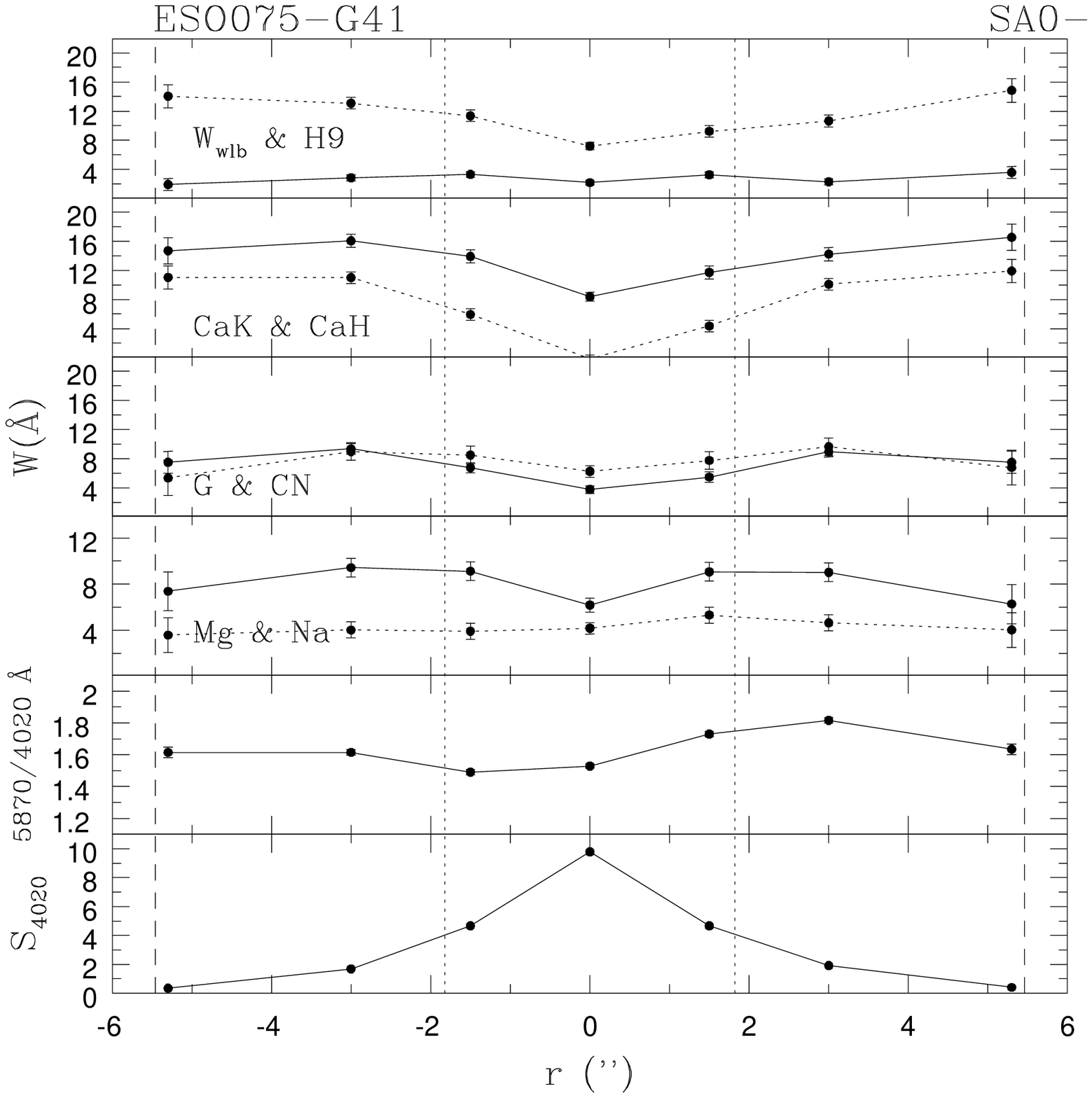}
\includegraphics{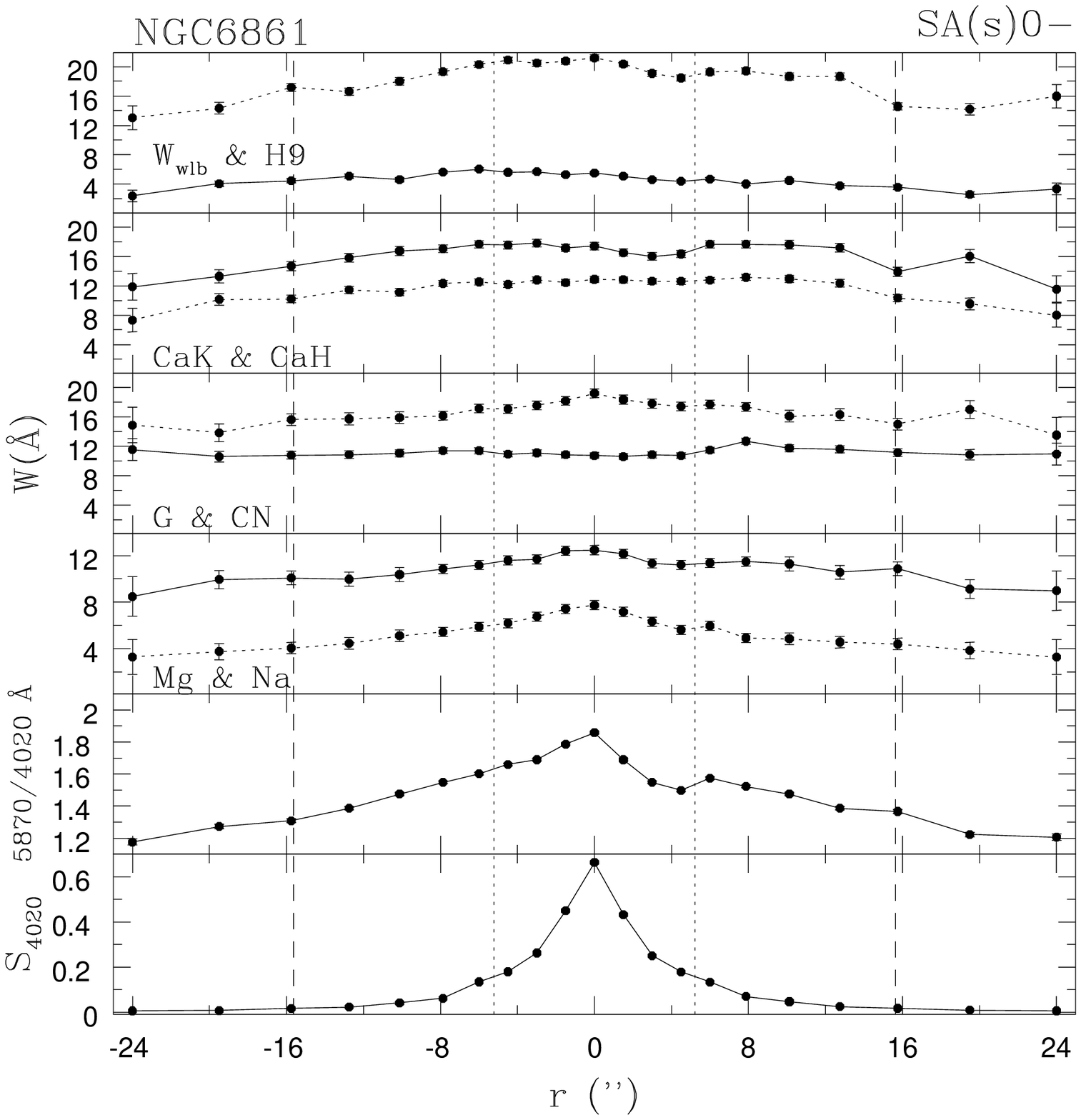}
\includegraphics{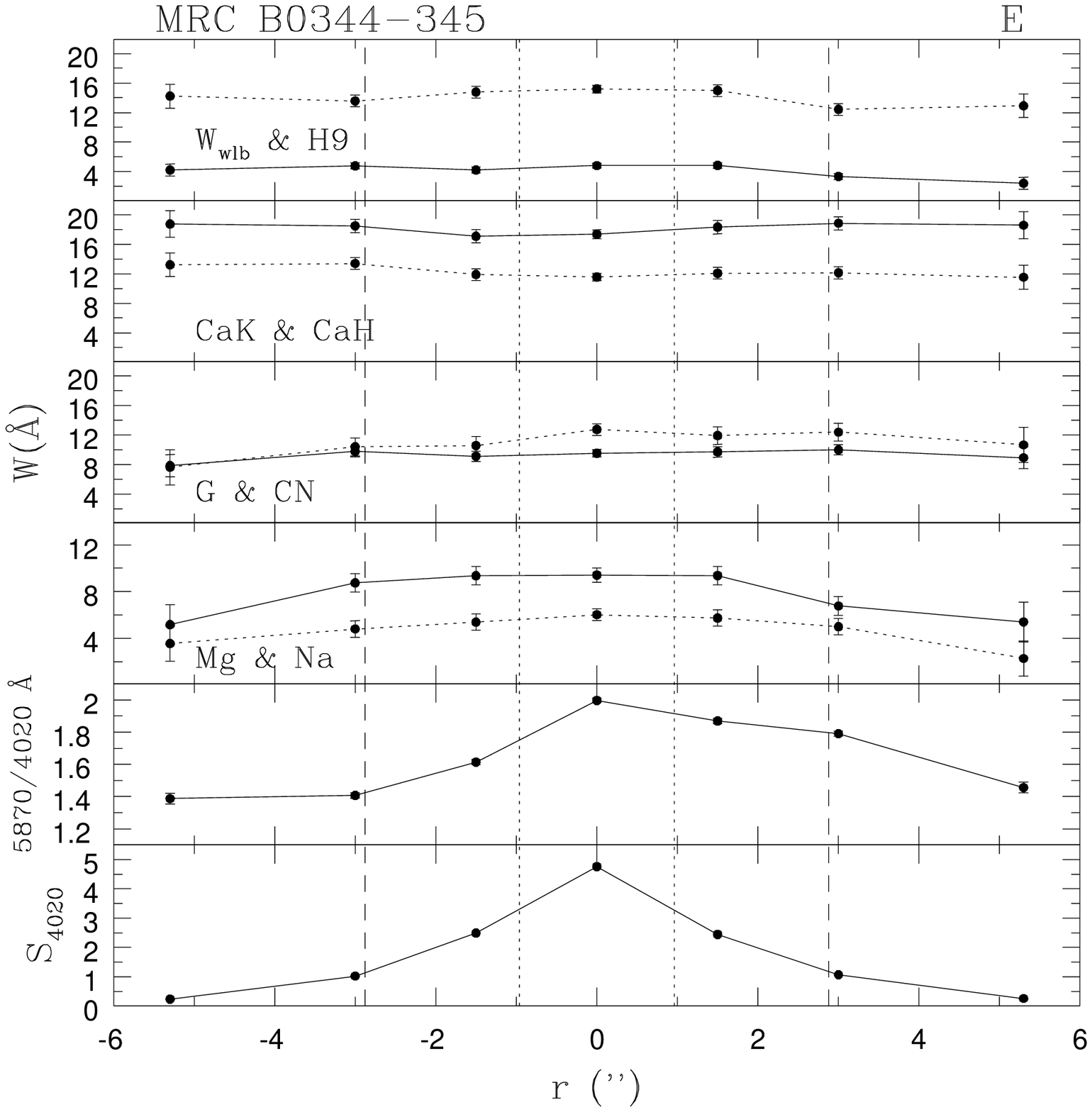}
\includegraphics{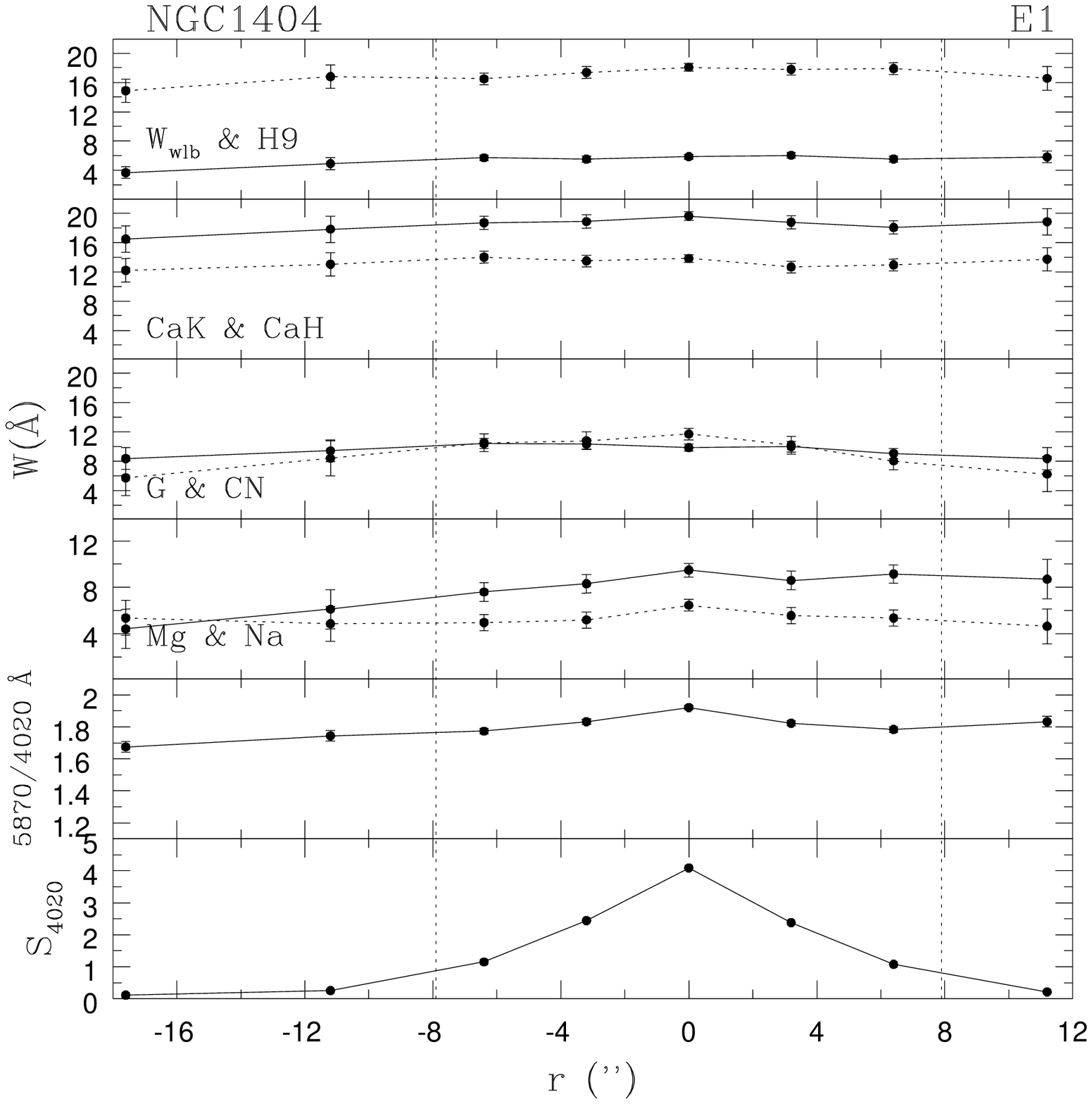}
\end{figure*}

\subsection{Radial variations of equivalent widths and continuum colours}

The variation of absorption line equivalent widths and continuum colours
as a function of distance from the nucleus allows us to study stellar
population gradients.  In non-active galaxies, the equivalent widths
usually increase from the external regions towards the bulge, where they
remain approximately constant. The presence of a burst of star-formation
and/or a featureless AGN continuum will produce a ``dilution'' of the
absorption lines, with a consequent decrease in equivalent width at the
nucleus in comparison with values at adjacent locations (Cid Fernandes et
al.\ 1998).

In Fig.\ \ref{variation} we illustrate the radial variations of
$W_\lambda$, the continuum flux ratio $F_{5870}$/$F_{4020}$ and the
surface brightness at 4020\,\AA\ for two radio galaxies and two control
sample galaxies of matched Hubble types. The dotted and dashed vertical
lines mark distances at each galaxy of 1\,kpc and 3\,kpc from the
nucleus, respectively. The radio galaxy ESO\,075-G41 shows dilution in
most equivalent widths and has a bluer continuum at the nucleus than
outside, while the radio galaxy MRC B0344$-$345 does not show dilution
and has a redder nucleus than its surroundings.  Some $W_\lambda$
profiles show a decrease with distance from the nucleus.  This latter behaviour
is observed also in the control sample galaxies.

In our radio galaxy sample, only the two BLRGs, ESO\,075-G41 and
Pictor\,A show clear dilutions of the nuclear equivalent widths and have
a bluer nuclear continuum when compared with the extranuclear spectra.  
This dilution is thus probably due to the presence of a featureless
continuum (hereafter called FC) being directly observed in these
galaxies. Most of the other radio galaxies do not show significant
variation in $W_\lambda$ along the slit, apart from a weak trend of
decreasing $W_\lambda$ away from the nucleus. The continuum is generally
redder close to the nucleus than outside.

The non-active galaxies of the control sample also show little variation
in equivalent width along the slit. Three of the lenticular galaxies from
the control sample show stronger gradients, with nuclear equivalent
widths $\sim 2$\AA\ larger than at 1\,kpc from the nucleus. In order to
investigate whether the gradients were due to the proximity of these
galaxies, thereby providing better spatial sampling, we binned several
extracted spatial elements into one, in order to sample the same spatial
extent as in the radio galaxies. Even after the binning, the gradients
were still present. We thus concluded that the gradients are probably
enhanced by the presence of the disk component in the lenticular
galaxies. Regarding the continuum, only the control galaxy NGC\,2865 has
a nuclear continuum bluer than in the extranuclear spectra; the others
all have redder nuclear continua.



\section{The spectral synthesis}

Spectral synthesis was performed using the probabilistic formalism
described in Cid Fernandes et al.\ (2001). We reproduced the observed
$W_\lambda$ and continuum ratios ($C_\lambda$) using a base of star
cluster spectra with different ages and metallicities (Bica \& Alloin
1986). We used 12 components representing the age-metallicity plane plus
a 13$^{th}$ component -- the FC component -- representing a canonical AGN
continuum $F_{\nu} \propto \nu^{-1.5}$ (Schmitt et al.\ 1999).

To synthesize the data from the two samples we used the continuum ratios
$C_{3660} = F(3660)/F(4020)$, $C_{4510} = F(4510)/F(4020)$, $C_{5870} =
F(5870)/F(4020)$, and $C_{6630} = F(6630)/F(4020)$, and the equivalent
widths $W_\mathrm{WLB}$, $W_\mathrm{H9}$, $W_\mathrm{Ca\,{\small II}~K}$,
$W_\mathrm{CN\,band}$, $W_\mathrm{G\,band}$ and $W_\mathrm{Mg\,{\small
I}+MgH}$. The adopted errors were $\sigma(W_\lambda)=0.4$\,\AA\ for
$W_\mathrm{Mg\,{\small I}+MgH}$, $\sigma(W_\lambda)=0.5$\,\AA\ for
$W_\mathrm{WLB}$, $W_\mathrm{H9}$, $W_{Ca\,{\small II}~K}$ and
$W_\mathrm{G\,band}$, 1.0\AA\ for W$_\mathrm{CN\,band}$ and
$\sigma(C_\lambda)=0.05$ for the continuum ratios (Cid Fernandes et al.\
1998). In a few cases the synthesis was performed with a smaller number
of equivalent widths, due to contamination from emission lines.

As pointed out by Storchi-Bergmann et al.\ (2000; see also Cid Fernandes
et al.\ 2001), it is not possible to discriminate the FC component from
the 3\,Myr young stellar component in this spectral range, for flux
contributions smaller than 40\,\% at 4020\,\AA , because they have very
similar continua. Therefore, in the description and discussion of the
synthesis results we have combined the 3\,Myr and FC components, which we
refer to as the 3\,Myr/FC component.

According to Cid Fernandes et al.\ (2001), this method of spectral
synthesis can have difficulty in accurately determining the contributions
of all 12 components of Bica's database when the S/N ratio is modest or
there is a reduced number of observables.  These constraints act
primarily in the sense of spreading a strong contribution in one
component preferentially among base elements of different metallicities
but of the same age.  Therefore, in order to produce more robust results,
we have grouped components of different metallicities but of the same age
into one component, characterized by that age. We have thus neglected the
potential differentiation in metallicity and concentrated on the more
robust age information.

In Tables \ref{resultsrg} and \ref{resultscontrol}, we present the
synthesis results as the relative contributions from components of four
age bins to the total flux at 4020\,\AA : 10\,Gyr, 1\,Gyr, 100$+$10\,Myr
and 3\,Myr/FC, in the nuclear region, at 1\,kpc, and at 3\,kpc from the
nucleus, respectively.

\begin{table*}
\caption{Percentage contributions of four age bins to the total flux at
4020\,\AA\ for
the radio galaxy sample, separated according to their FR classification.}
\label{resultsrg}
\begin{center}
\begin{tabular}{lccccccccccccc} \hline
Name     & \multicolumn{3}{c}{10\,Gyr} & \multicolumn{3}{c}{1\,Gyr} &
\multicolumn{3}{c}{100$+$10\,Myr}  & \multicolumn{3}{c}{3\,Myr/FC} & Mean Age \\
         & Nuc. & 1\,kpc & 3\,kpc & Nuc. & 1\,kpc & 3\,kpc & Nuc.
& 1\,kpc & 3\,kpc & Nuc. & 1\,kpc & 3\,kpc & (Gyr) \\ \hline
{\it FR I} \\
ESO\,350-G15  & 56 & 57   & 55 & 41 & 41   & 39 & 3 & 2  & 5 & 0 & 0 & 1 & 3.4 \\
ESO\,198-G1   & 53 & 54   & 51 & 44 & 43   & 46 & 3 & 3  & 2 & 0 & 0 & 1 & 3.2 \\
MRC B0332$-$391   & 64 & 61   & 56 & 30 & 33   & 34 & 5 & 5  & 9 & 1 & 1 & 1 & 3.2 \\
IC\,2082      & 60 & 54   & 65 & 38 & 44   & 33 & 2 & 2  & 2 & 0 & 0 & 0 & 3.8 \\
MRC B0429$-$616   & 50 & 50   & 53 & 42 & 43   & 39 & 7 & 6  & 7 & 1 & 1 & 1 & 2.2 \\
MRC B0620$-$526   & 47 & 46   & 45 & 46 & 48   & 54 & 6 & 5  & 1 & 1 & 1 & 0 & 2.1 \\
ESO\,161-IG7  & 59 & 58   & 53 & 38 & 38   & 43 & 3 & 3  & 4 & 0 & 1 & 0 & 3.6 \\
MRC B0715$-$362   & 72 & 61   & -- & 17 & 33   & -- & 9 & 5  & -- & 2 & 1 & -- & 2.8 \\
ESO\,377-G46  & 62 & 59   & -- & 35 & 40   & -- & 2 & 1  & -- & 1 & 0 & -- & 3.2 \\
MRC B2013$-$557   & 49 & 46   & 33 & 42 & 46   & 61 & 7 & 6  & 4 & 2 & 1  & 2 & 1.7 \\
MRC B2148$-$555   & 62 & 64   & 61 & 35 & 33   & 33 & 3 & 3  & 4 & 0 & 0  & 2 & 3.9 \\
ESO\,349-G10  & 61 & 62   & 59 & 35 & 33   & 36 & 3 & 4  & 4 & 1 & 1  & 1 & 3.1 \\
\\
{\it FR II} \\
NGC\,612      & 68 & 67   & 60 & 21 & 22   & 16 & 9 & 9  & 21 & 2 & 2 & 2 & 2.6 \\
ESO\,248-G10  & 52 & 47   & 47 & 44 & 49   & 47 & 3 & 3  & 5 & 1 & 1  & 1 & 2.5 \\
Pictor\,A    & 21(45) & 49   & -- & 10(14) & 8   & -- & 12(41) & 8  & -- & 57 & 35 & -- & (1.1) \\
ESO\,365-IG6  & 69 & 68   & -- & 28 & 29   & -- & 3 & 3  & -- & 0 & 0 & -- & 4.6 \\
MRC B1413$-$364   & 63 & 52   & -- & 33 & 43   & -- & 3 & 4  & -- & 1 & 1 & -- & 3.2 \\
MRC B1637$-$771   & 46 & 50   & 48 & 46 & 43   & 44 & 7 & 6  & 7 & 1 & 1  & 1 & 2.0 \\
ESO\,075-G41  & 39(55) & 46   & 80 & 10(13) & 23   & 11 & 24(32) & 15  & 7 & 27 & 16 & 2 & (1.7) \\
AM\,2158$-$380  & 69 & 74   & 70 & 25 & 19   & 22 & 4 & 5  & 6 & 2 & 2 & 2 & 3.0 \\
\\
{\it FR x} \\
MRC B0344$-$345   & 63 & 61   & 54 & 29 & 32   & 44 & 6 & 6  & 1 & 2 & 1 & 1 & 2.5 \\
MRC B0456$-$301   & 14 & 13   & -- & 60 & 51   & -- & 24 & 31  & -- & 2 & 5 & -- & 0.5 \\
ESO\,271-G20  & 61 & 54   & 56 & 36 & 44   & 41 & 2 & 2  & 2 & 1 & 0 & 0 & 3.2 \\
ESO\,338-IG11 & 49 & 50   & 51 & 48 & 48   & 47 & 2 & 2  & 2 & 1 & 0 & 0 & 2.4 \\ \hline
\end{tabular}
\end{center}
\end{table*}

\begin{table*}
\caption{Percentage contributions of four age bins to the total flux
at 4020\,\AA\ for
elliptical and lenticular galaxies from the control sample.}
\label{resultscontrol}
\begin{center}
\begin{tabular}{lcccccccccccc} \hline
Name     & \multicolumn{3}{c}{10\,Gyr} & \multicolumn{3}{c}{1\,Gyr} &
\multicolumn{3}{c}{100$+$10\,Myr} & \multicolumn{3}{c}{3\,Myr/FC} \\
         & Nuc. & 1\,kpc & 3\,kpc & Nuc. & 1\,kpc & 3\,kpc & Nuc.
& 1\,kpc & 3\,kpc & Nuc. & 1\,kpc & 3\,kpc \\ \hline
{\it Ellipticals}&&&&&&&&&&&&\\
NGC\,1404  & 75 & 79   & -- & 22 & 16   & -- & 3 & 5  & -- & 0 & 0  & -- \\
NGC\,1700  & 71 & 71   & -- & 22 & 25   & -- & 6 & 3  & -- & 1 & 1  & -- \\
NGC\,2865  & 56 & 65   & -- & 24 & 19   & -- & 20 & 15  & -- & 0 & 1  & -- \\
NGC\,3091  & 76 & 74   & -- & 15 & 22   & -- & 8 & 4  & -- & 1 & 0  & -- \\
NGC\,3585  & 73 & 82   & -- & 20 & 14   & -- & 6 & 4  & -- & 1 & 0  & -- \\
NGC\,3904  & 81 & 78   & -- & 15 & 20   & -- & 3 & 2  & -- & 1 & 0  & -- \\
NGC\,3923  & 69 & 77   & -- & 25 & 19   & -- & 6 & 4  & -- & 0 & 0  & -- \\
NGC\,4936$^1$& 75 & 88   & -- & 19 & 10   & -- & 4 & 2  & -- & 2 & 0 & -- \\
NGC\,4936$^2$& 81 & 80   & -- & 11 & 16   & -- & 6 & 3  & -- & 2 & 1 & -- \\
NGC\,5061  & 70 & 77   & -- & 18 & 16   & -- & 12 & 6  & -- & 0 & 1  & -- \\
NGC\,5328  & 66 & 67   & -- & 24 & 18   & -- & 9 & 14  & -- & 1 & 1  & -- \\
NGC\,5813  & 83 & --   & -- & 8 & --   & -- & 7 & --  & -- & 2 & --  & -- \\
{\it Lenticulars}&&&&&&&&&&&&\\
NGC\,3706  & 78 & 76 & -- & 12 & 19 & -- & 8 & 4 & -- & 2 & 1 & -- \\
NGC\,4373  & 66 & 68 & -- & 30 & 26 & -- & 4 & 5 & -- & 0 & 1 & -- \\
NGC\,4825  & 80 & 81 & 79 & 12 & 12 & 11 & 7 & 6 & 6 & 1 & 1 & 3 \\
NGC\,6684  & 66 & 54 & -- & 32 & 43 & -- & 2 & 3 & -- & 0 & 0 & -- \\
NGC\,6861  & 67 & 53 & 37 & 27 & 44 & 53 & 6 & 3 & 10 & 0 & 0 & 0 \\
NGC\,7049  & 73 & 56 & 34 & 21 & 42 & 62 & 6 & 2 & 3 & 0 & 0 & 1 \\
NGC\,7079  & 50 & 45 & -- & 48 & 52 & -- & 2 & 3 & -- & 0 & 0 & -- \\
\hline
$^1$ 1.5m ESO\\
$^2$ 3.6 NTT
\end{tabular}
\end{center}
\end{table*}

\subsection{Synthesis results for the radio galaxies}

The synthesis results for the radio galaxies are summarized in Table
\ref{resultsrg}. In most of the radio galaxies, the nuclear and
extranuclear stellar population is dominated by old (10\,Gyr) and
intermediate (1\,Gyr) age components. 

There are only four radio galaxies
in which the younger or power-law components contribute more than
10\,\% of the total flux at 4020\,\AA, either at the nucleus or outside.
Two of the latter galaxies are the BLRGs Pictor\,A and ESO\,075-G41, in
which the 3\,Myr/FC component is probably dominated by the FC, as
discussed in previous sections. 

For the two BLRG's galaxies, we performed
another synthesis after subtracting the contribution of
the 3\,Myr/FC component, in order to test whether the underlying stellar
population was similar to that in the other galaxies. We show the results of 
this new synthesis in parenthesis in Table \ref{resultsrg}:
the contribution of the 10\,Gyr stellar population increases,
indeed becoming more similar to that of the other galaxies,
but the contribution of the 1\,Gyr component does not,
thus maintaining the difference from the other galaxies. There is
an increase of the contribution of the 100+10\,Myr component
in the new synthesis. Finally, we point out that,
for these galaxies, the synthesis results must
be taken with caution, due to the fact that
we have assumed a fixed slope for the power-law component (namely
$F_\nu \propto \nu^{-1.5}$).
As this component is very strong in these galaxies,
if the real slope is different, it will
affect the  results of the synthesis.
In particular, if the slope is harder than assumed, 
we may see excess blue light which could appear artificially
as a large contribution of the 100+10\,Myr component.

\smallskip \noindent{\it The 10\,Gyr component --} In 21 of the 24 radio 
galaxies the 10\,Gyr component contributes from $\sim$50\,\% to 70\,\% 
of the total flux in the nuclear region. The contribution is smaller only
in MRC\,B0456-301 and in the two BLRG´s before the subtraction of the
3\.Myr/FC component. There is little variation with radius. 
Only five objects have differences between nuclear and extranuclear
contributions larger than 10\,\%. In three cases the contribution of
the 10\,Gyr component decreases outwards and in the two BLRGs it increases, as
expected. After subtraction of the 3\,Myr/FC component from the spectra
of the two BLRGs, the gradient disappears in Pictor\,A and is weakened in
ESO\,075-G41.

\smallskip \noindent {\it The 1\, Gyr component --} The 1\,Gyr component
contribution is dominant only in the nuclear spectrum of MRC B0456$-$301, and
it contributes at least $\sim$30\,\% of the total flux in 19 objects. In four
objects the difference between nuclear and extranuclear contributions is
$\ge 10$\,\%, with the contribution of this component increasing
outwards.

\smallskip \noindent {\it Younger components --} Besides the
two BLRG´s, the FR\,x galaxy MRC\,B0456-301 is the only radio-galaxy
which presents, at the nucleus and up to 1\,kpc from it,
a contribution of the 100+10\,Myr component significantly larger than
10\,\%. NGC\,612 displays such contribution at 3\,kpc from the nucleus, probably
triggered by an interaction related to its peculiar morphology. 
After subtracting the contribution of the 3\,Myr/FC component 
from the spectra of Pictor\,A and ESO\,075-G41, there is an increase
of the contribution of the 100+10\,Myr component,

\smallskip \noindent {\it Internal reddening --} Internal reddening in
the radio galaxies is generally small.  The only cases of significant
reddening, with values in the range $0.06\le E(B-V)_\mathrm{int}\le 0.4$,
are observed in the FR\,II/x galaxies NGC\,612, ESO\,248-G10,
ESO\,075-G41 and MRC B0344$-$345, and in the FR\,I galaxy MRC B2013$-$557.

\smallskip \noindent {\it Gradients --} The 3\,Myr/FC component decreases
outwards in the two BLRGs, but, as discussed above, this is most probably
due to the unresolved FC component. By examining a stellar spatial
profile, we concluded that at the angular distance corresponding at the
galaxies to 1\,kpc, there is still contamination of the spectra by a
possible point source at the nucleus. This is the reason why, at 1\,kpc,
Table \ref{resultsrg} shows some contribution from the 3\,Myr/FC
component. A true stellar population gradient is observed in the FR\,II
galaxy NGC\,612, in which the stellar population is predominantly old at
the nucleus and there is a 100+10\,Myr-old component at a radius of 
3\,kpc.

%
%
%
%
%
%


\subsection{Synthesis results for the control sample galaxies}

The synthesis results for the control sample galaxies
are summarized in Table \ref{resultscontrol}.

\smallskip\noindent {\it The 10\,Gyr component --} The nuclear and
extranuclear stellar populations of the non-active galaxies are dominated
by the old (10\,Gyr) component, which in most cases contributes 70 to
85\,\% of the total flux at 4020\,\AA.

\smallskip \noindent {\it The 1\,Gyr component --} The intermediate age
(1\,Gyr) component is also significant in this sample and its
contribution in most cases is within the range 15 to 30\,\%.

\smallskip \noindent {\it Younger age components --} There are three
non-active galaxies in which the 100+10\,Myr component contributes 10\,\%
or more of total flux at 4020\,\AA, but this contribution is never larger than 20\,\%.
These three galaxies are ellipticals. The 3\,Myr/FC component
contribution is not significant in any of the galaxies of the control
sample.

\smallskip \noindent {\it Reddening --} None of the control sample
galaxies shows internal reddening larger than $E(B-V)_\mathrm{int}=0.05$.

\smallskip \noindent {\it Gradients --} In general we do not observe any
population gradients in the galaxies of the control sample. Only three
lenticular galaxies show differences between the nuclear and extranuclear
age components at a level greater than 10\,\%. In these three cases, the
10\,Gyr component contribution decreases while the 1\,Gyr contribution
increases outwards. We attribute these gradients to the presence of a
disk component in these galaxies, as discussed in Section 3.2.

\begin{figure*}
\vspace{12cm}
\caption{Histograms for the radio galaxies showing the contributions of
different age components (10\,Gyr, 1\,Gyr, 100+10\,Myr and 3\,Myr/FC) to
the total flux at 4020\,\AA.  FR\,I galaxies are shown as open
histograms with a heavy outline and FR\,II/x are shown hatched.}
\label{histo_radiofr1xfr2}
\includegraphics{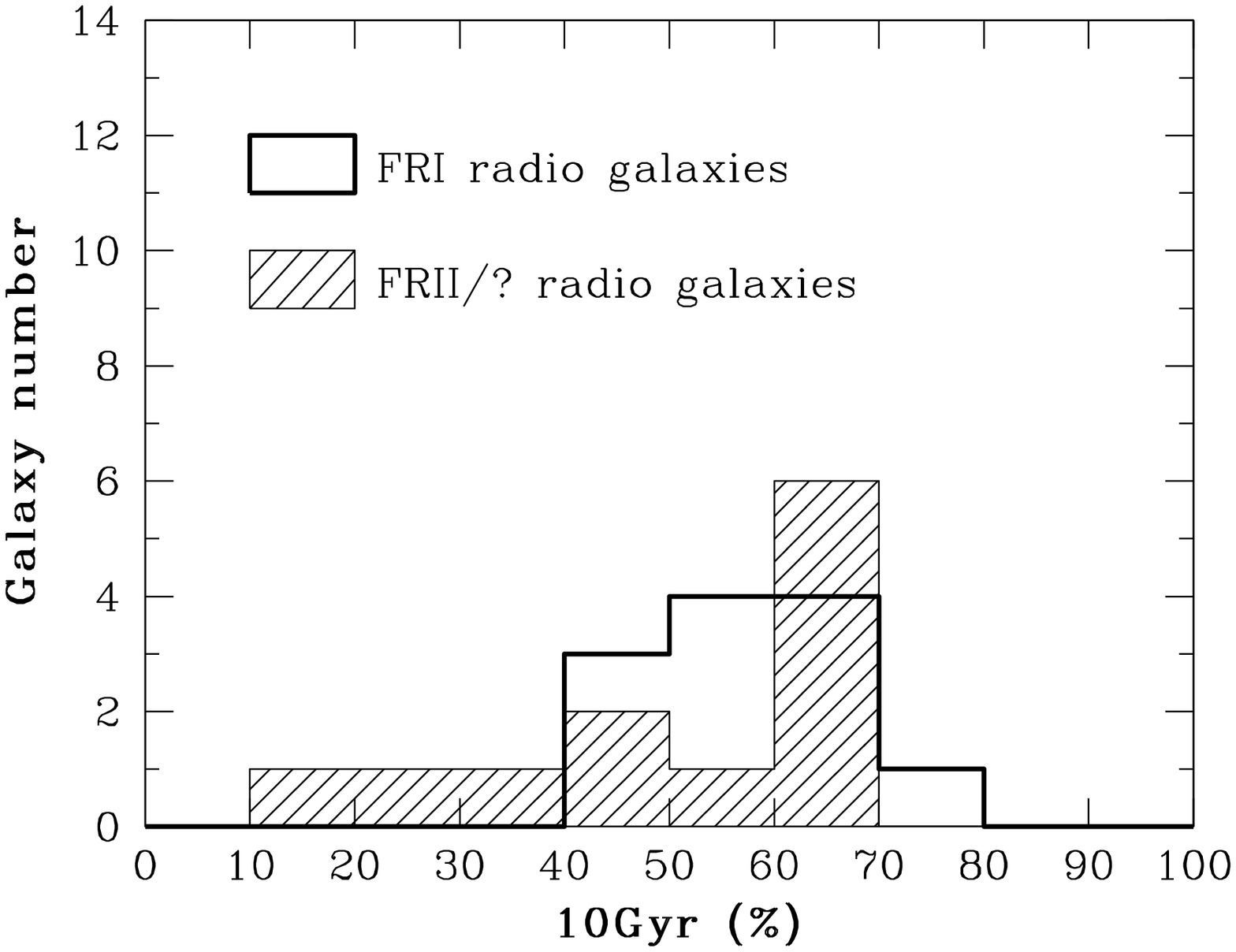}
\includegraphics{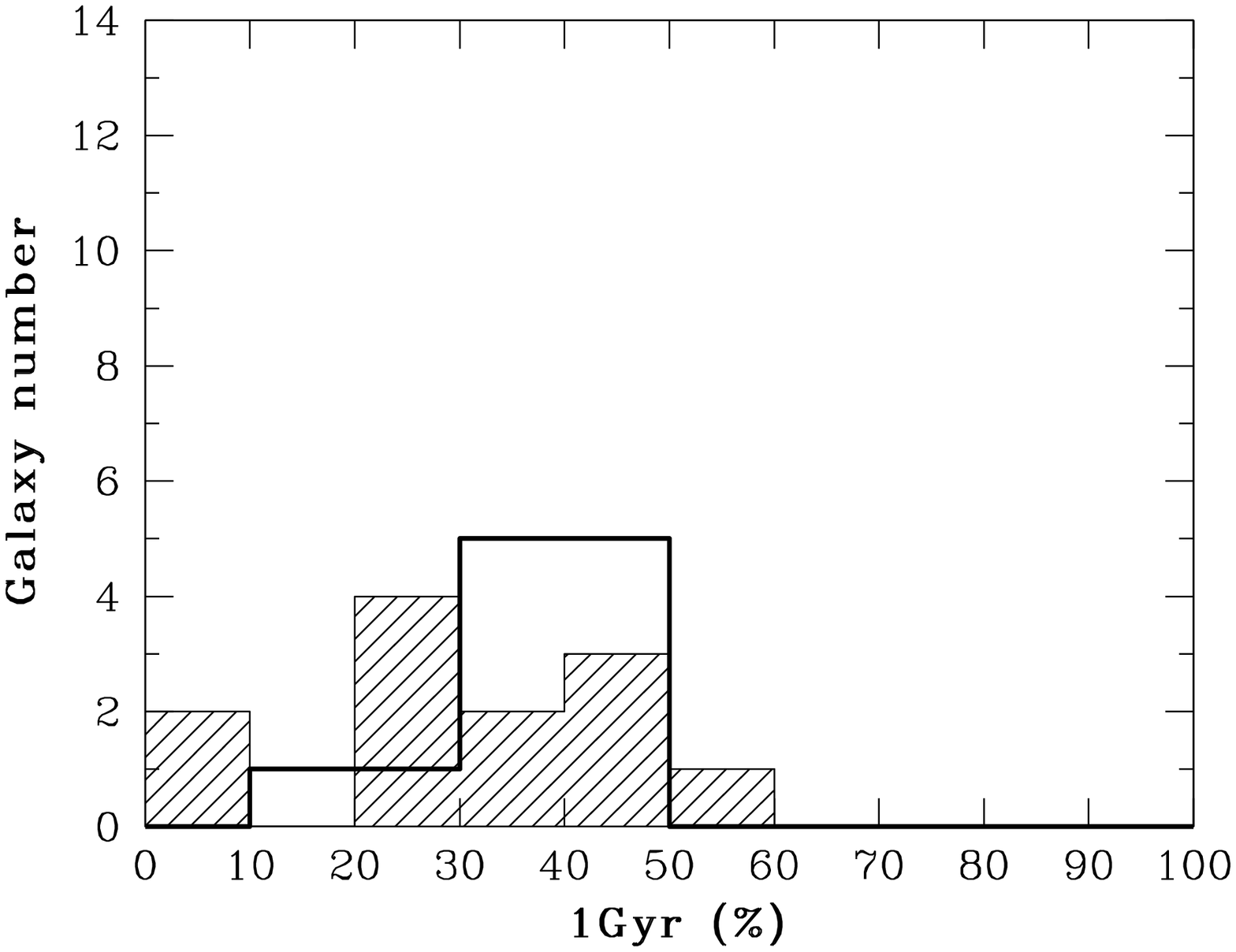}
\includegraphics{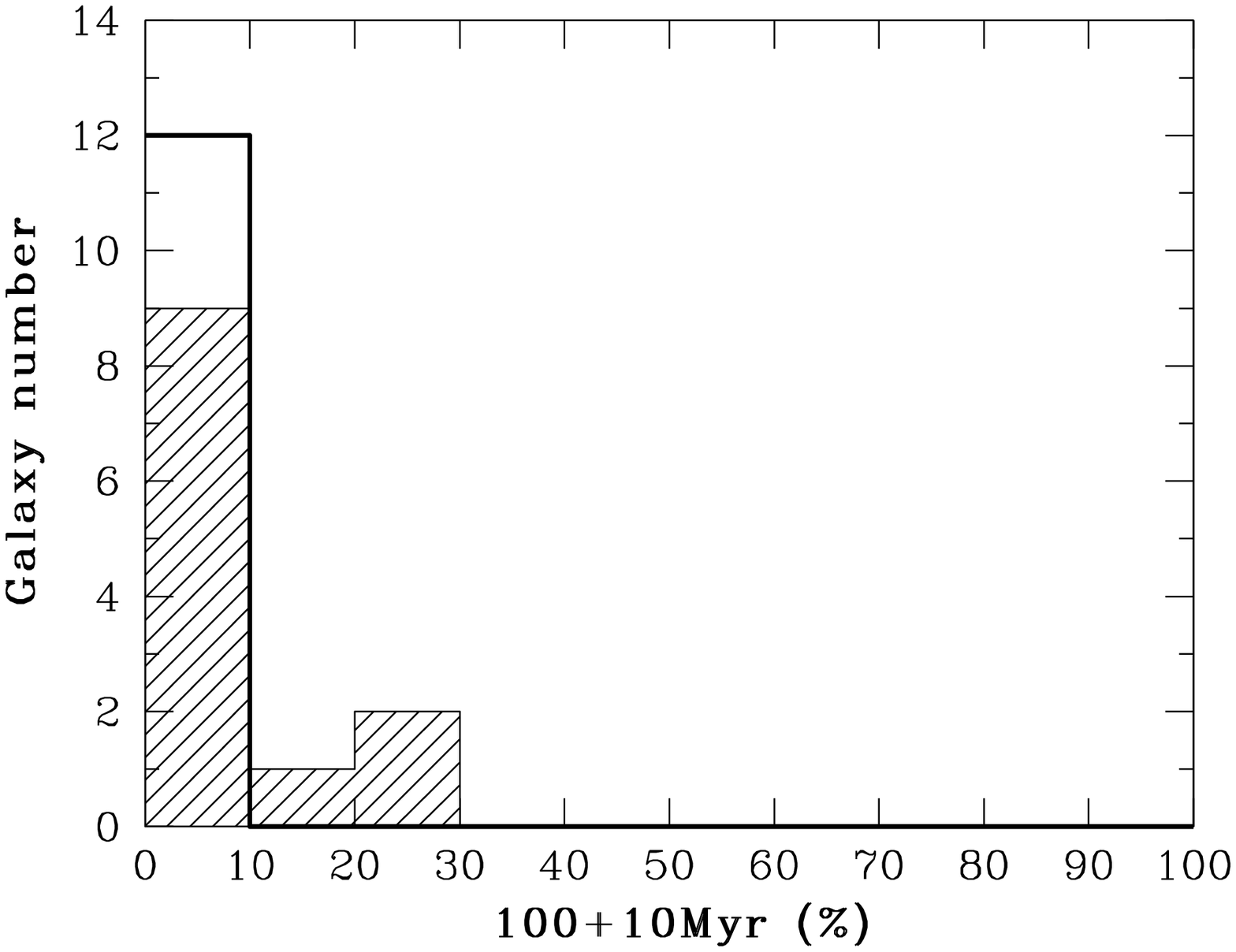}
\includegraphics{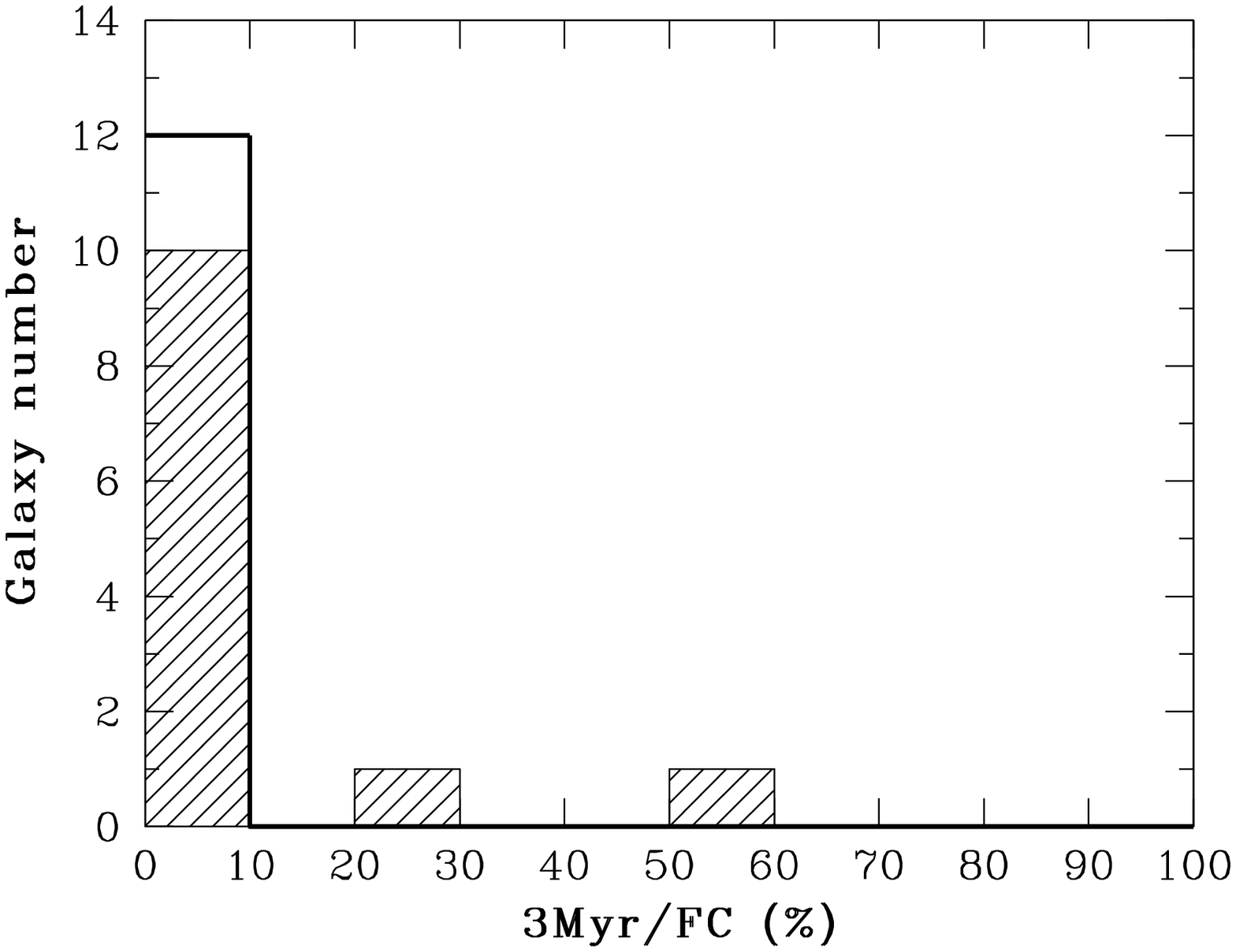}
\end{figure*}

\begin{table*}
\caption{The mean percent contribution (and corresponding standard
deviation) to the total flux at 4020\,\AA\ for the nuclear stellar
populations of the 10\,Gyr, 1\,Gyr, 100+10\,Myr and 3\,Myr/FC components.}
\label{res_sin_radio}
\begin{center}
\begin{tabular}{lllll} \hline
Objects                 & 10\,Gyr     & 1\,Gyr      & 100+10\,Myr &  3\,Myr/FC \\ \hline
FR\,I radio galaxies    & 57.9  (6.9) & 36.9  (7.4) & 4.4 (2.2) & 0.8 (0.7) \\
FR\,II/x radio galaxies & 51.2 (17.7) & 32.5 (14.6) & 8.3  (7.6) & 8.1 (16.4) \\ \hline
All radio galaxies      & 54.5 (13.9) & 34.7 (11.8) & 6.3  (5.9) & 4.4 (12.2) \\
Control sample          & 65.4 (17.3) & 24.9 (11.1) & 5.4  (4.1) & 0.6 (0.6) \\ \hline
\end{tabular}
\end{center}
\end{table*}
\section{Discussion}

\subsection{Comparison FR\,I vs. FR\,II galaxies}

In Fig.\ \ref{histo_radiofr1xfr2} we compare the population synthesis
results of the FR\,I galaxies with those of the FR\,II, plus intermediate
and uncertain types, for each age component.
In each histogram, we show the number of galaxies as a function of the
percentage contribution of that age component to the total flux at
4020\,\AA. We show the results only for the nucleus as we did not find
significant spatial variation, as discussed in the previous sections.
In Table \ref{res_sin_radio} we list the mean percent contribution of
each age bin to the total flux at 4020\,\AA\ and the corresponding
standard deviations.

The histograms show that the fractional contributions of each age
component have a narrower distribution in FR\,I than in FR\,II galaxies.
In other words, the stellar populations in FR\,I galaxies are more
homogeneous than in FR\,II galaxies. No FR\,I galaxy has more than a
10\,\% contribution from components of age 100\,Myr or younger, while
four FR\,II/x galaxies have such components.

The above results translate into mean contributions (Table
\ref{res_sin_radio}) of the 10\,Gyr and 1\,Gyr age components slightly
larger in FR\,I than in FR\,II galaxies, while the reverse is true for
the 100\,Myr and younger components. The standard deviations are larger
for FR\,II galaxies in accordance with their broader distributions in 
Fig.\ \ref{histo_radiofr1xfr2}.


\begin{figure*}
\vspace{12cm}
\caption{As in Fig. \ref{histo_radiofr1xfr2} for the radio galaxies identified
according to
their emission-line spectra: BLRG, NLRG, WLRG and NO-E.}
\label{emissao}
\includegraphics{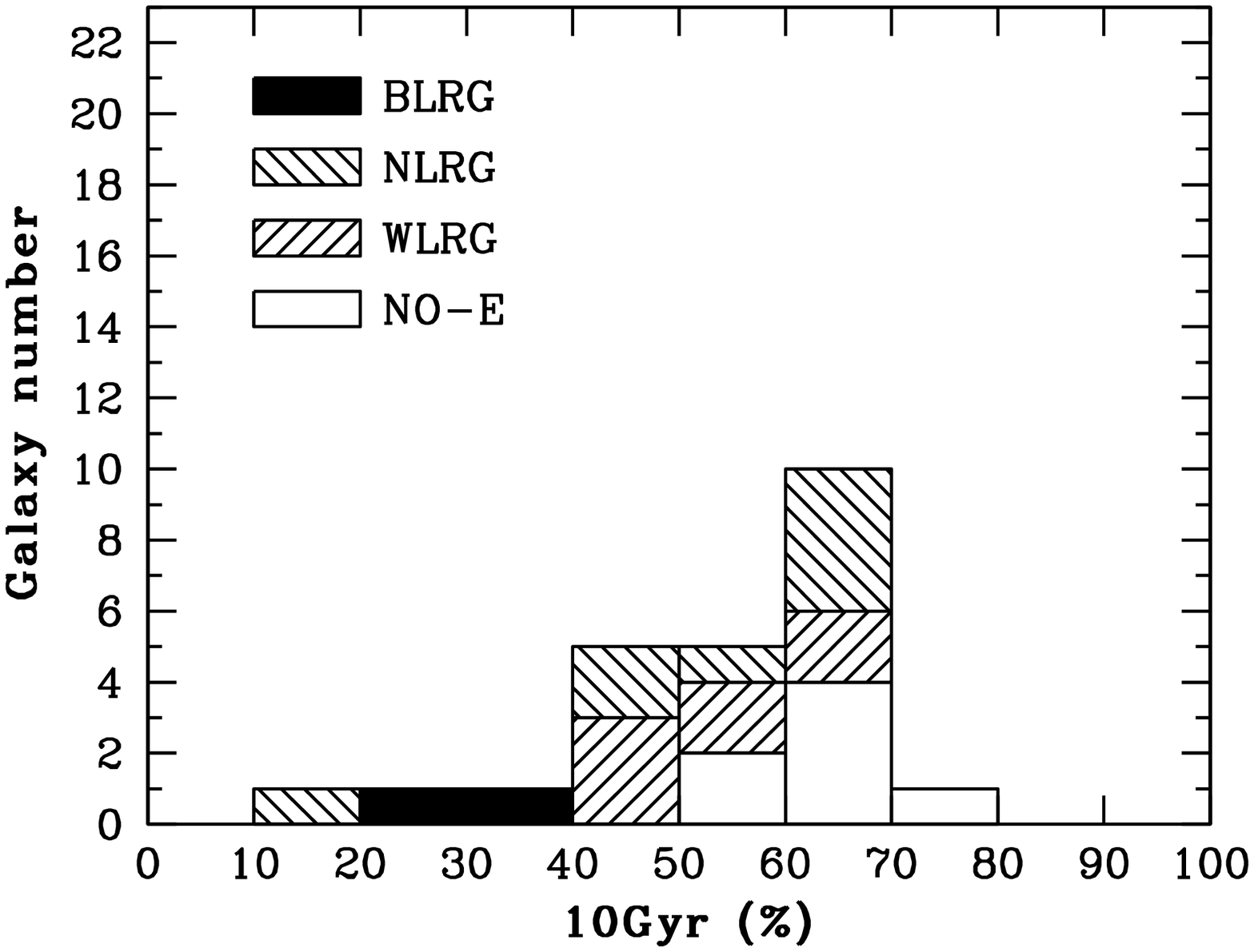}
\includegraphics{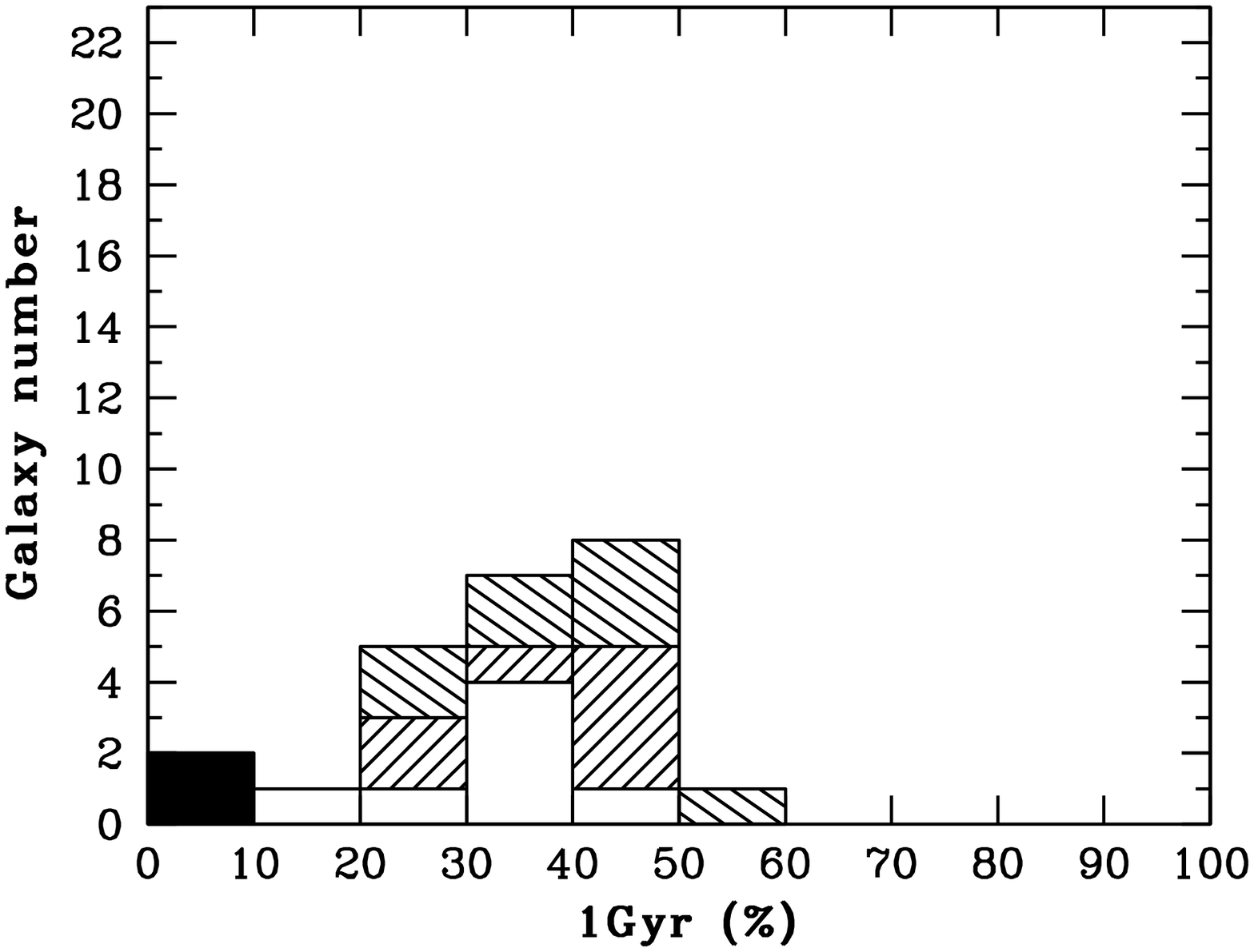}
\includegraphics{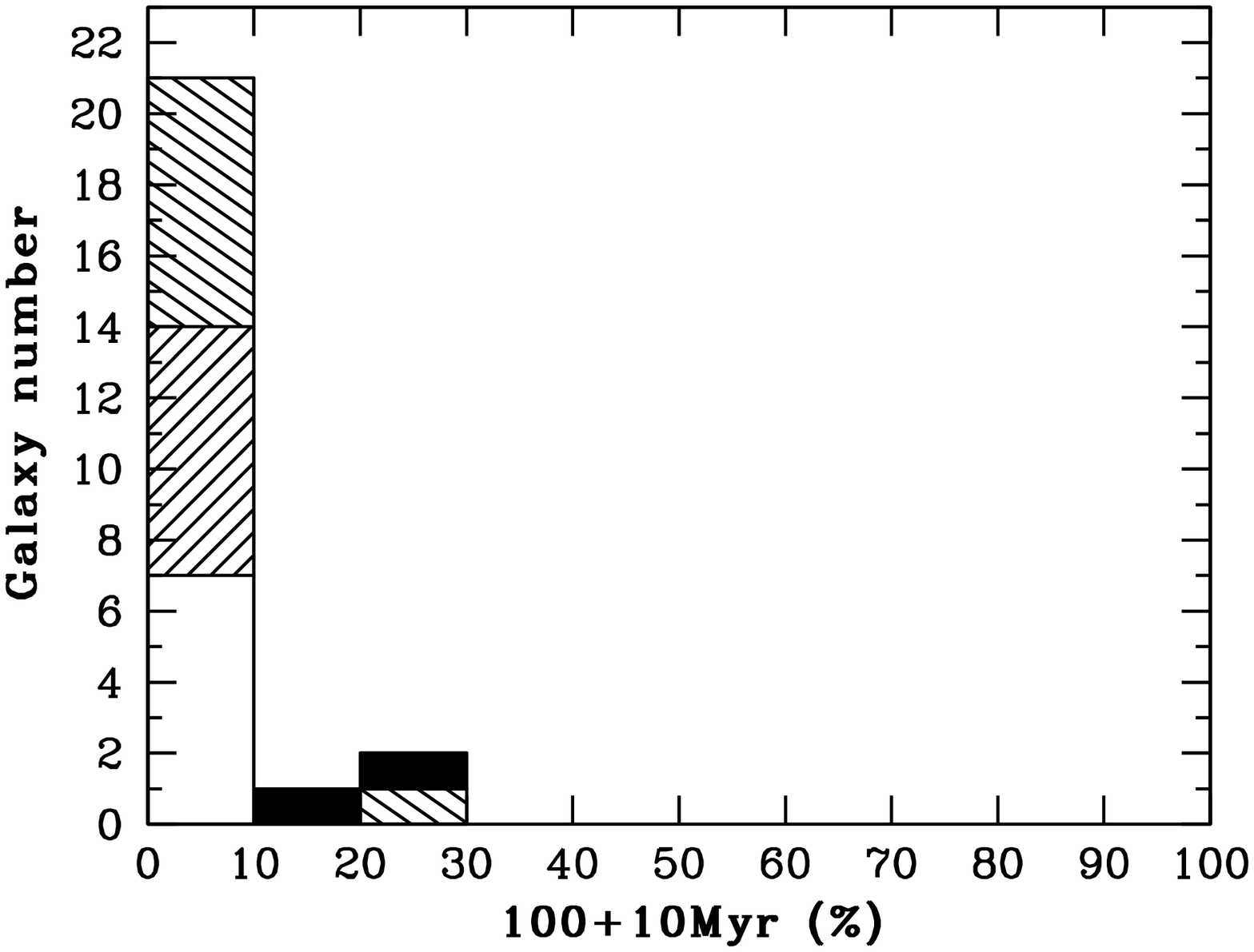}
\includegraphics{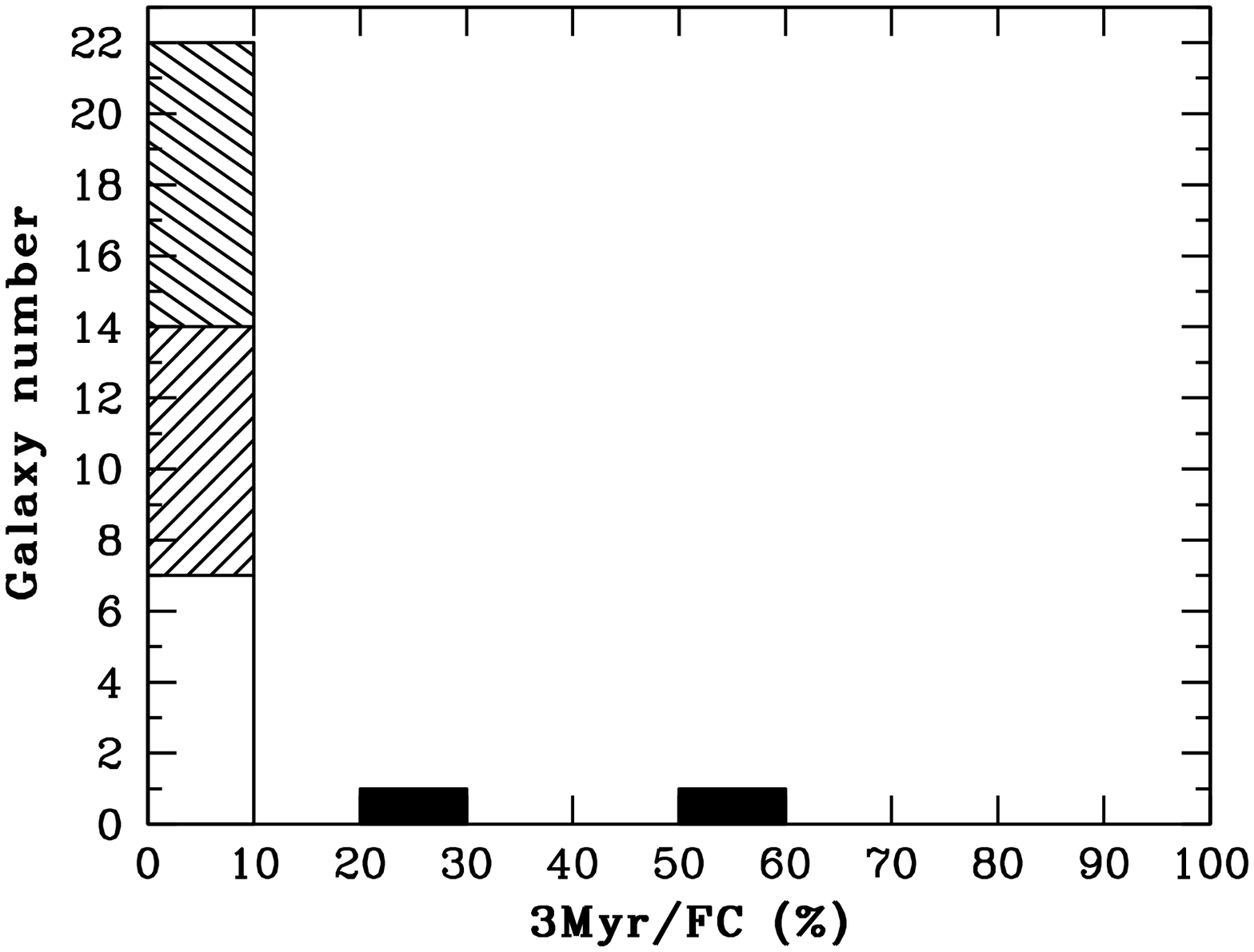}
\end{figure*}

\subsection{Relation with Emission Line Properties}

Fig.\ \ref{emissao} shows the distribution of fractional contributions of
the four age components to the spectra of the radio galaxies where they
are identified according to their emission-line spectra (BLRG, NLRG, WLRG
and NO-E). The NO-E radio galaxies tend to have the largest contribution
from the 10\,Gyr stellar component ($>50$\,\%), similar to that of the
control sample (see section 5.4). The smallest contribution ($<40$\,\%)
of the 10\,Gyr component is observed in the two BLRGs and in the NLRG
MRC\,B0456$-$301. Most of the NLRGs and WLRGs show intermediate values
(40--70\,\%) for contributions of the 10\,Gyr component. Regarding the
1\,Gyr component, the only clear trend is that the BLRGs have the
smallest contributions ($<10$\,\%). In the case of the $10+100$\,Myr age
component, the only three galaxies with a contribution larger than 10\,\%
are the two BLRGs and one NLRG (B\,0456$-$301); these are the same three
galaxies which show the smallest contribution of the 10\,Gyr component.  
In the case of the 3\,Myr/FC component, only the two BLRGs show
contributions larger than 10\,\%, which, as noted earlier, we attribute
mostly to the FC.

\begin{figure*}
\vspace{12cm}
\caption{As in Fig. \ref{histo_radiofr1xfr2} for the radio galaxies identified
as a function of radio power (in units of $10^{25}$ W\,Hz$^{-1}$).}
\label{luminosidade}
\includegraphics{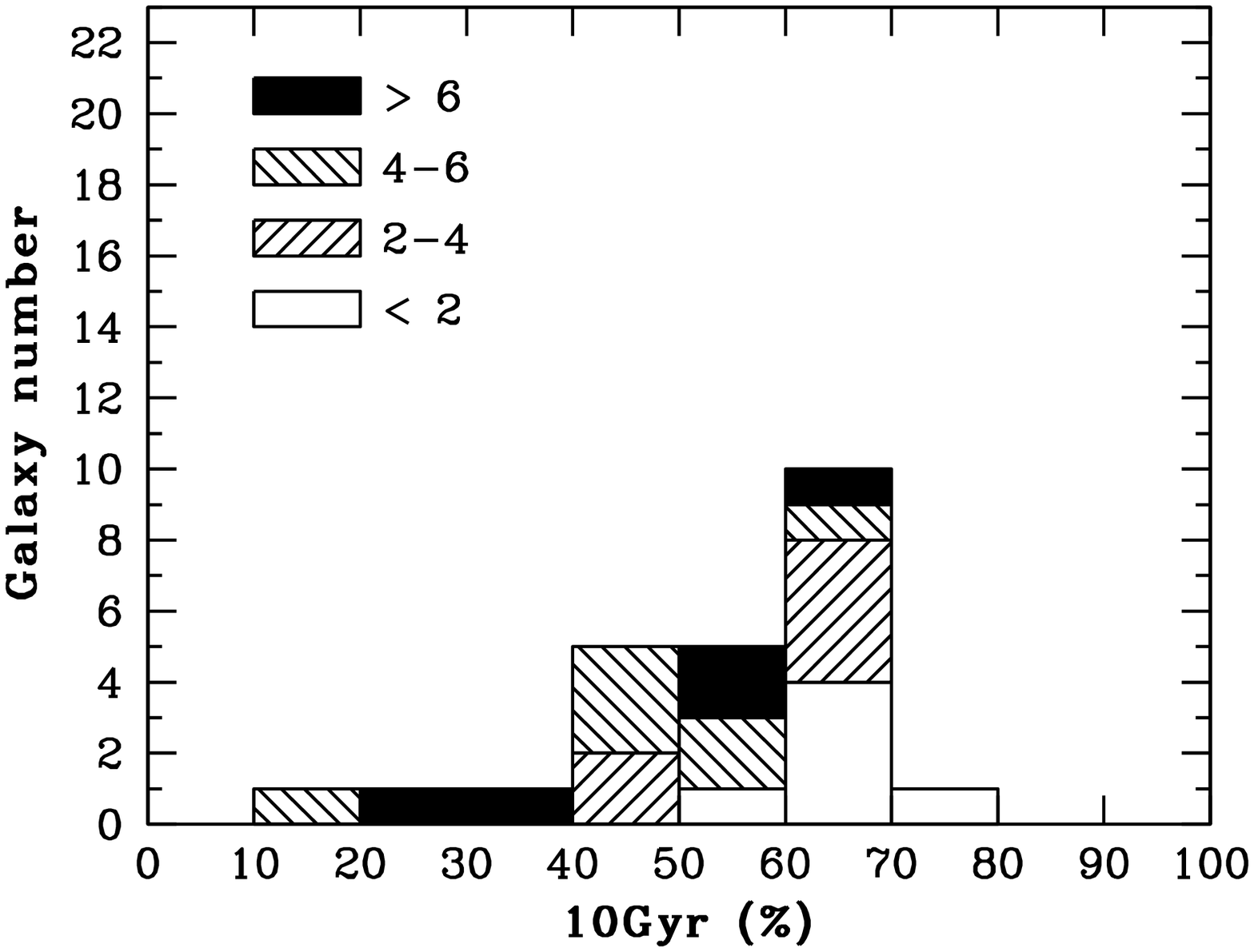}
\includegraphics{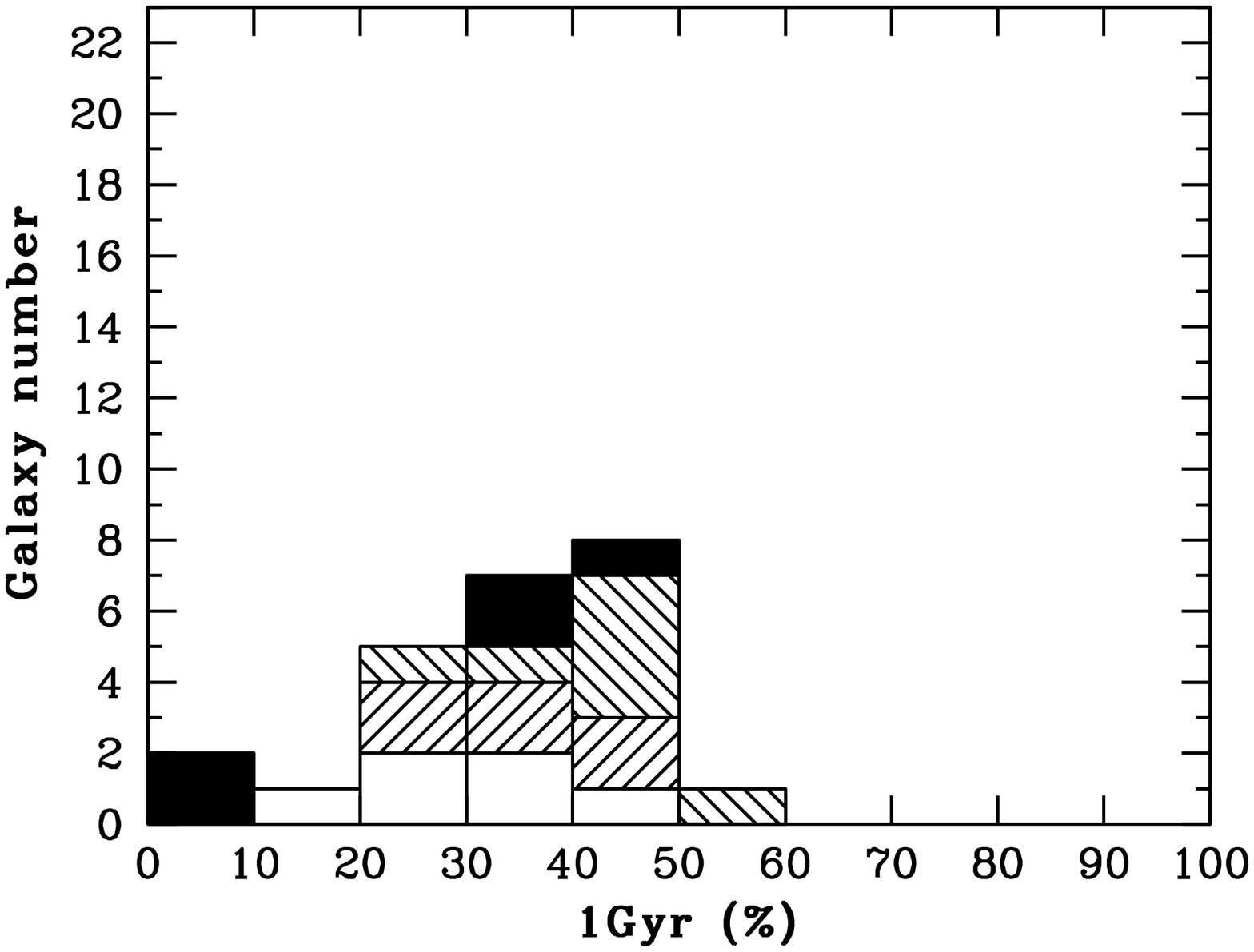}
\includegraphics{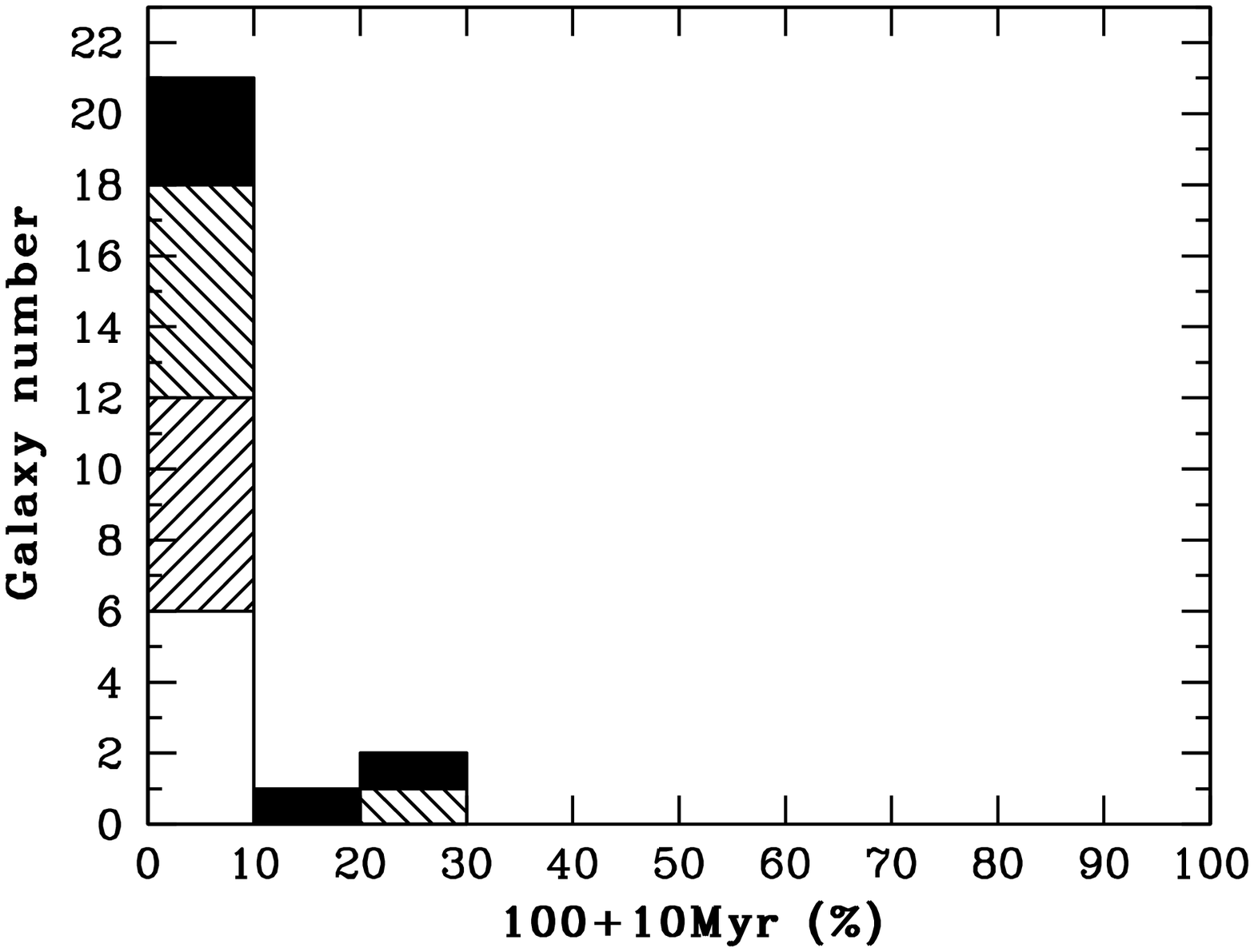}
\includegraphics{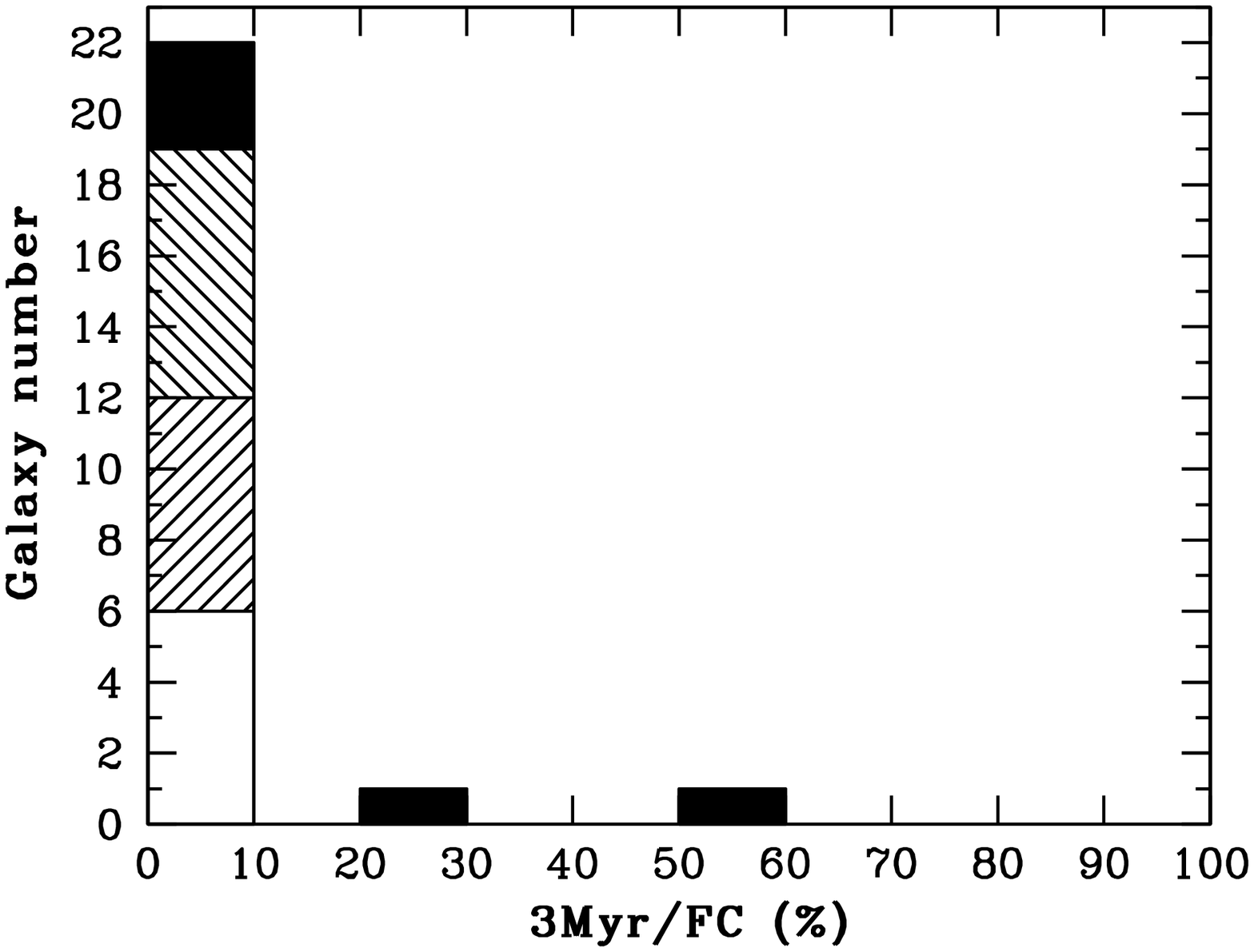}
\end{figure*}

In 13 radio galaxies we were able to measure the flux of the [O\,{\sc
iii}] $\lambda$5007 emission line.  In order to investigate the relation
between the luminosity $L_\mathrm{[O\,{\scriptsize III}]}$ in this line
and the age of the stellar population, we defined a percentage-weighted
``mean age'' $\bar t$ as the mean decimal logarithm of the stellar
population ages used in the synthesis, as follows:


\begin{equation}
\log(\bar t) \equiv \sum x_i log(t_i).
\end{equation}

\noindent For the four ages used here, this becomes (in yr):

\begin{equation}
\log(\bar t) \equiv 10x_{10}+9x_9+8x_8+7x_7
\end{equation}

\noindent where the $x_i$ are the fractional contributions to the total
flux at 4020\,\AA\ of the components with ages $t_i=10^{10}$, $10^9$, $10^8$ and $10^7$ yr.
The values of $\bar t$ are listed in the last column of Table \ref{resultsrg}. For
the two BLRGs, we calculated $\bar t$ excluding the 3\,Myr/FC component.

\begin{figure}
\vspace{8cm}
\caption{Relation between the luminosity $\log L_\mathrm{[O\,{\scriptsize III}]}$
and the mean age, as defined in the text.}
\label{l[OIII]}
\includegraphics{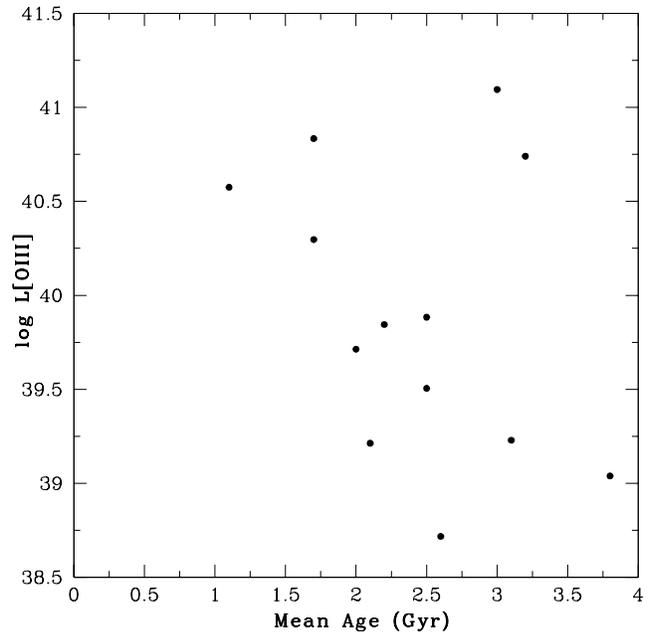}
\end{figure}

The relation between $\log(L_\mathrm{[O\,{\footnotesize III}]}$) and
$\log(\bar t)$ can be seen in Fig.\ \ref{l[OIII]} for the nuclear
spectra.  Although the age spread is not large, there is a tendency for
younger galaxies to have higher emission-line luminosities, while
older galaxies seem to have a larger spread in their emission-line
luminosities.

\subsection{Relation with Radio Power}

Although the radio luminosities do not cover a wide range, we also find
some correlation between the radio power and the stellar populations
properties of the radio galaxies. We have separated the radio sample into four power
bins: $\le$2, 2--4, 4--6 and $\ge 6\times 10^{25}$\,W\,Hz$^{-1}$.
The values of radio power are listed in the last column of Table 
\ref{amostrarg2}. The distribution of the
fractional contributions of the four age components split by radio power
is shown in Fig.\ \ref{luminosidade}. The galaxies with lowest radio
power tend to have the largest contribution from the 10\,Gyr age
component, and to have a stellar population mix closest to that of the
control sample (see section 5.4). For high radio power, the stellar
population properties vary widely.

\begin{figure}
\vspace{8cm}
\caption{Relation between the log radio power (in units of  W\,Hz$^{-1}$)
and the mean age, as defined in the text.}
\label{lradio}
\includegraphics{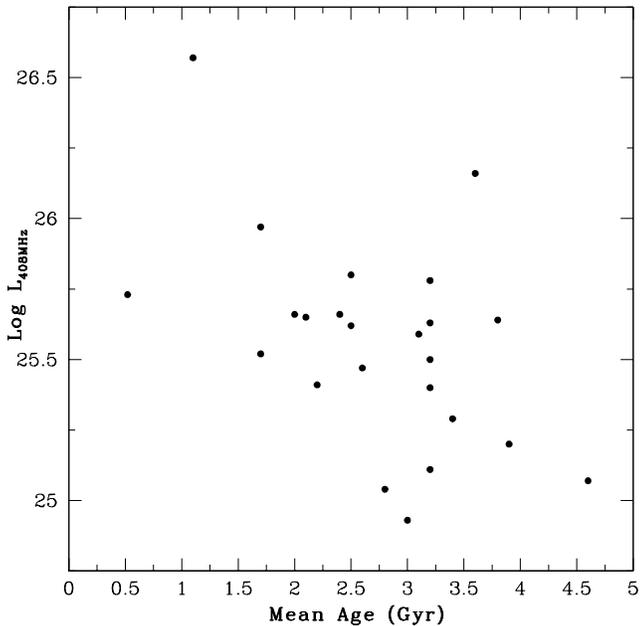}
\end{figure}

In order to better quantify the relation between the radio power and the
age of the stellar population, we have plotted in Fig.\ \ref{lradio} the
logarithm of the radio luminosity at 408\,MHz, log(L$_{408\,MHz}$)
against the mean age as calculated in the previous
section. We observe an inverse relation in Fig.\ \ref{lradio}: the most
powerful radio sources tend to be found in the youngest galaxies.

\begin{figure*}
\vspace{12cm}
\caption{Fractional contribution (in percent) of the different age components
plotted against log radio power (in units W\,Hz$^{-1}$).}
\label{lradio_age}
\includegraphics{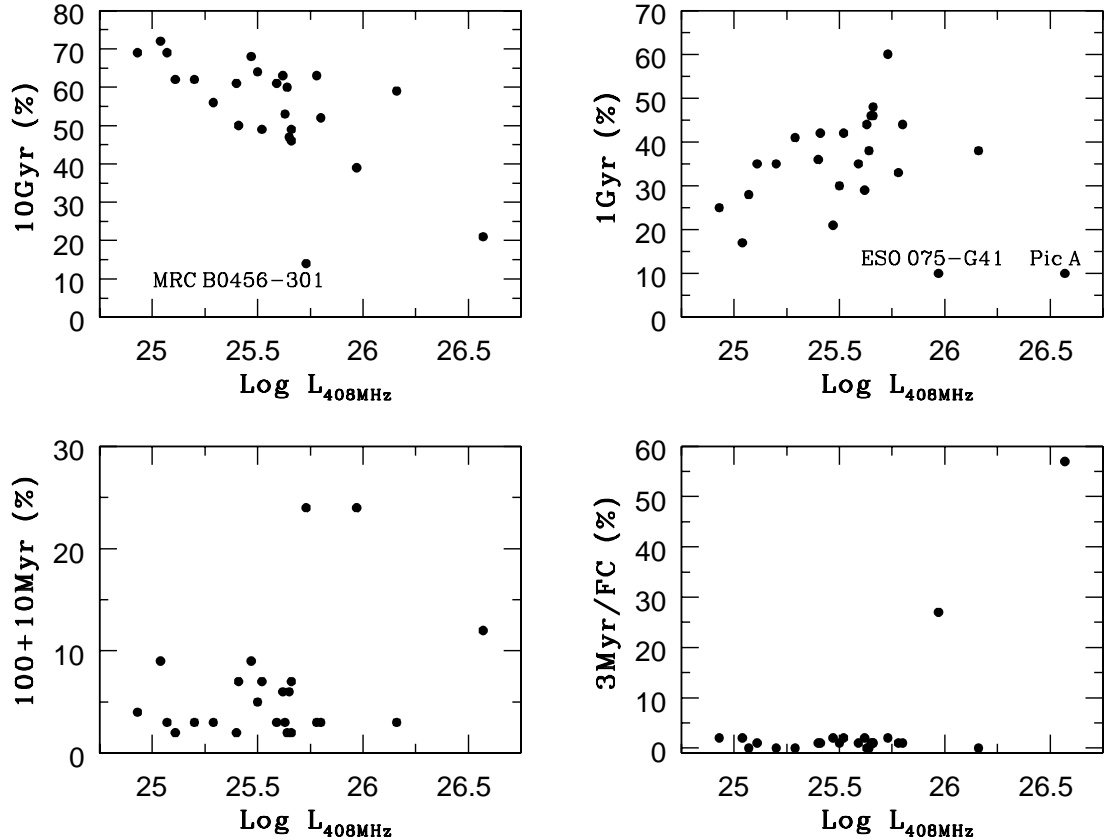}
\end{figure*}

In Fig.\ \ref{lradio_age} we show the percentage contribution of each age
bin to the total flux at $\lambda$4020 against log(L$_{408\,MHz}$).
For most of the sample, there is an inverse relation
between the contribution of the 10\,Gyr component and the radio power: the 
most radio luminous galaxies present the smallest contribution of the
10\,Gyr component.
Only one galaxy does not follow  the relation: MRC B\,0456-301, whose nuclear
light is dominated by the contribution of the 1\,Gyr and 100+10\,Myr components.  

There is also a direct relation between the contribution of the 1\,Gyr component
and radio power: larger radio power corresponds to larger contribution of
the 1\,Gyr component. The two BLRG´s seem to be exceptions to this relation.
Nevertheless, as pointed out in Section 4.1, the synthesis in these two cases
may be uncertain due to the large contribution of the direct AGN light
to the spectrum.

There is no obvious correlation of
radio power with the younger age components.

\subsection{Radio galaxies vs.\ control sample}

\begin{figure*}
\vspace{13cm}
\caption{
Histograms for the radio and control sample galaxies, comparing the
relative contributions of different age components to the total flux at
4020\,\AA.  Open histograms with a heavy outline show the non-active
galaxies while hatched histograms show the radio galaxies.}
\label{histo_radioxnaoativa_1.5kpc}
\includegraphics{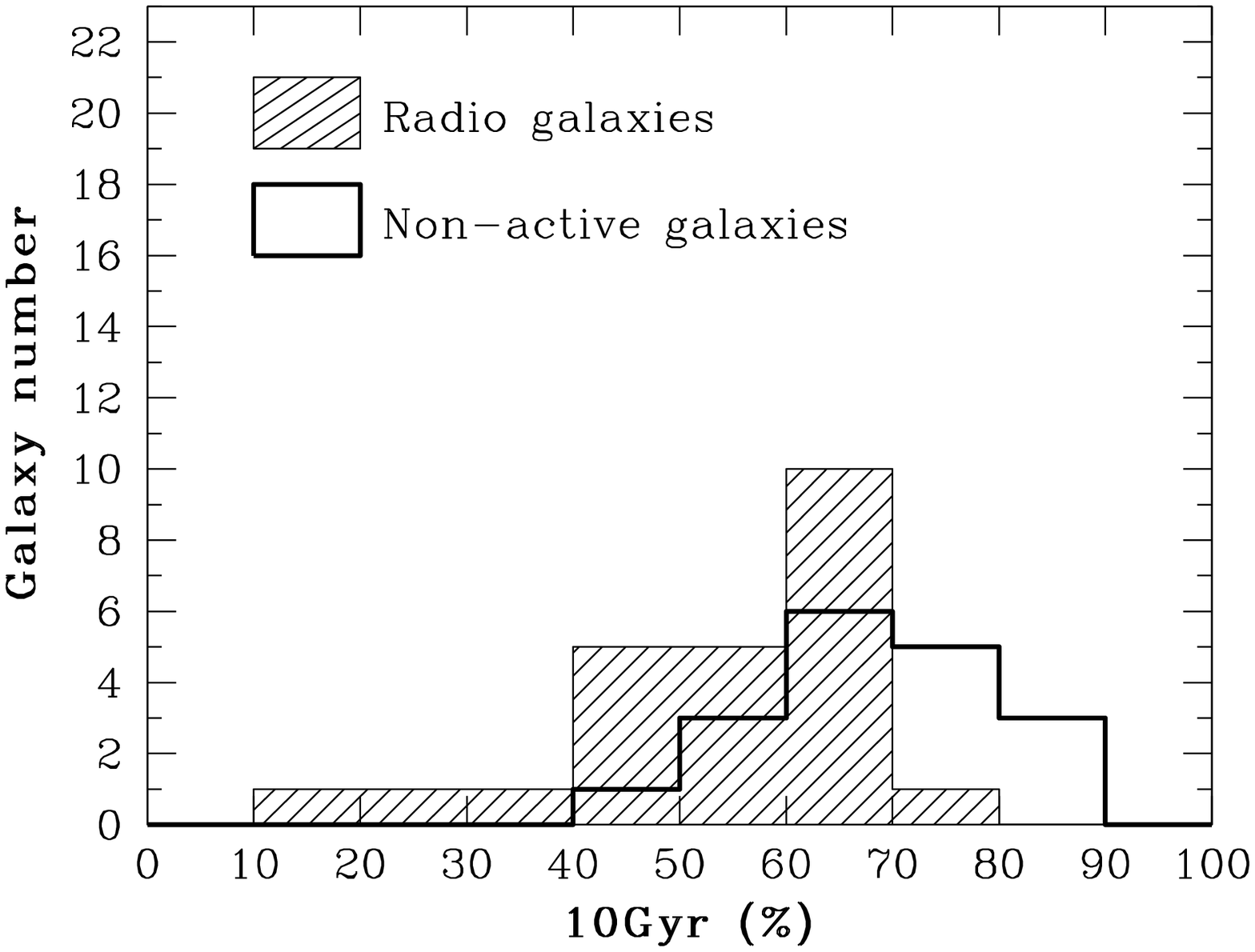}
\includegraphics{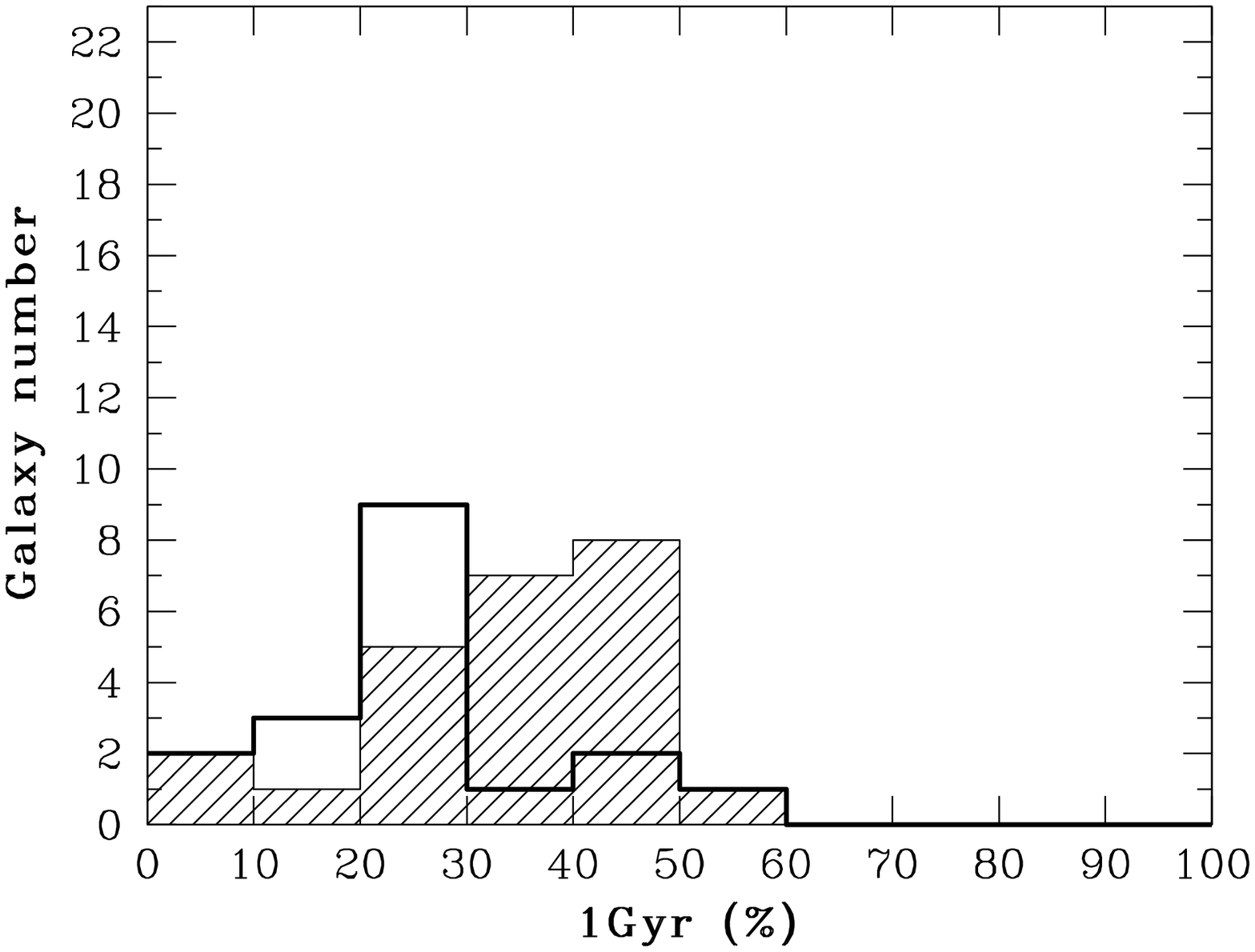}
\includegraphics{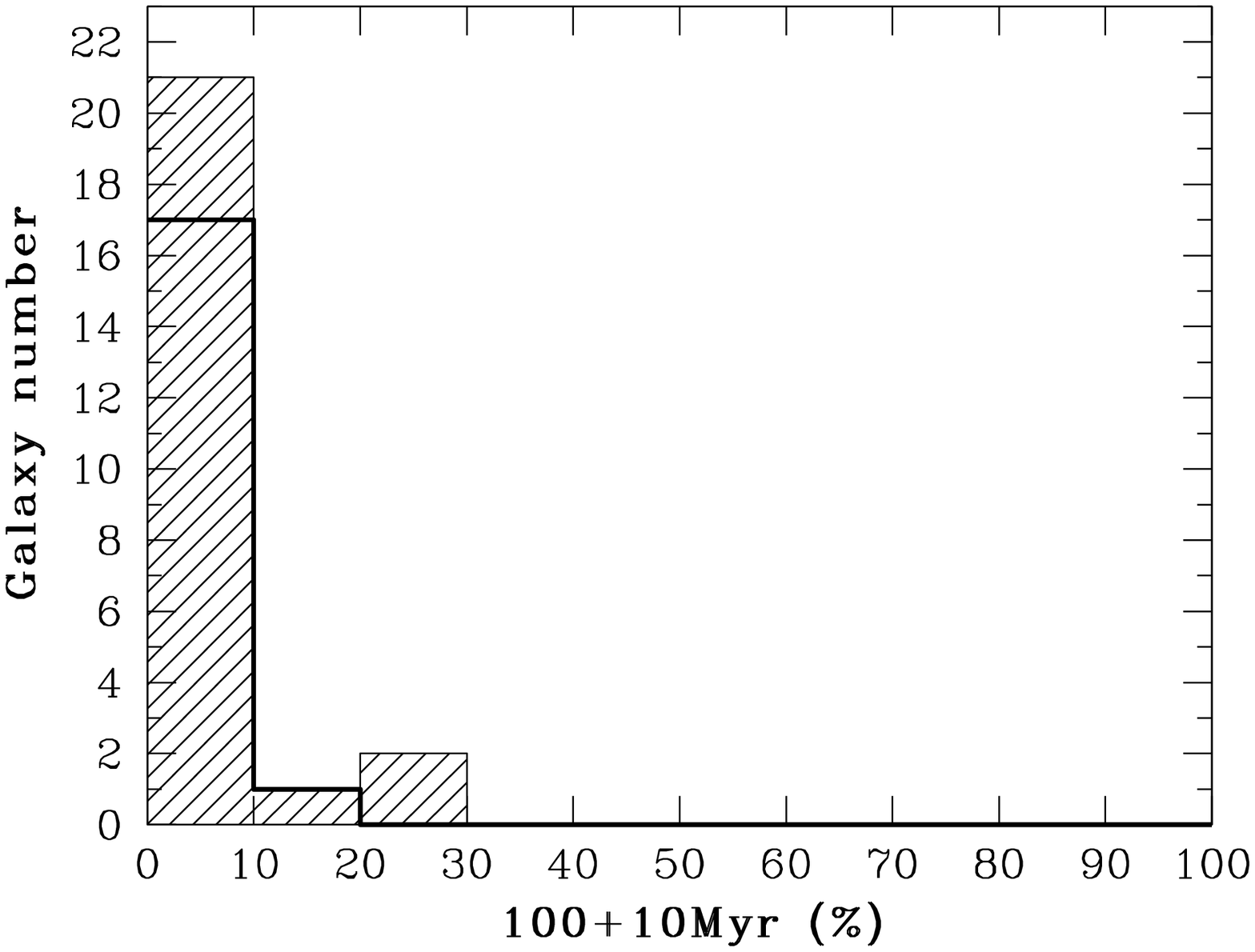}
\includegraphics{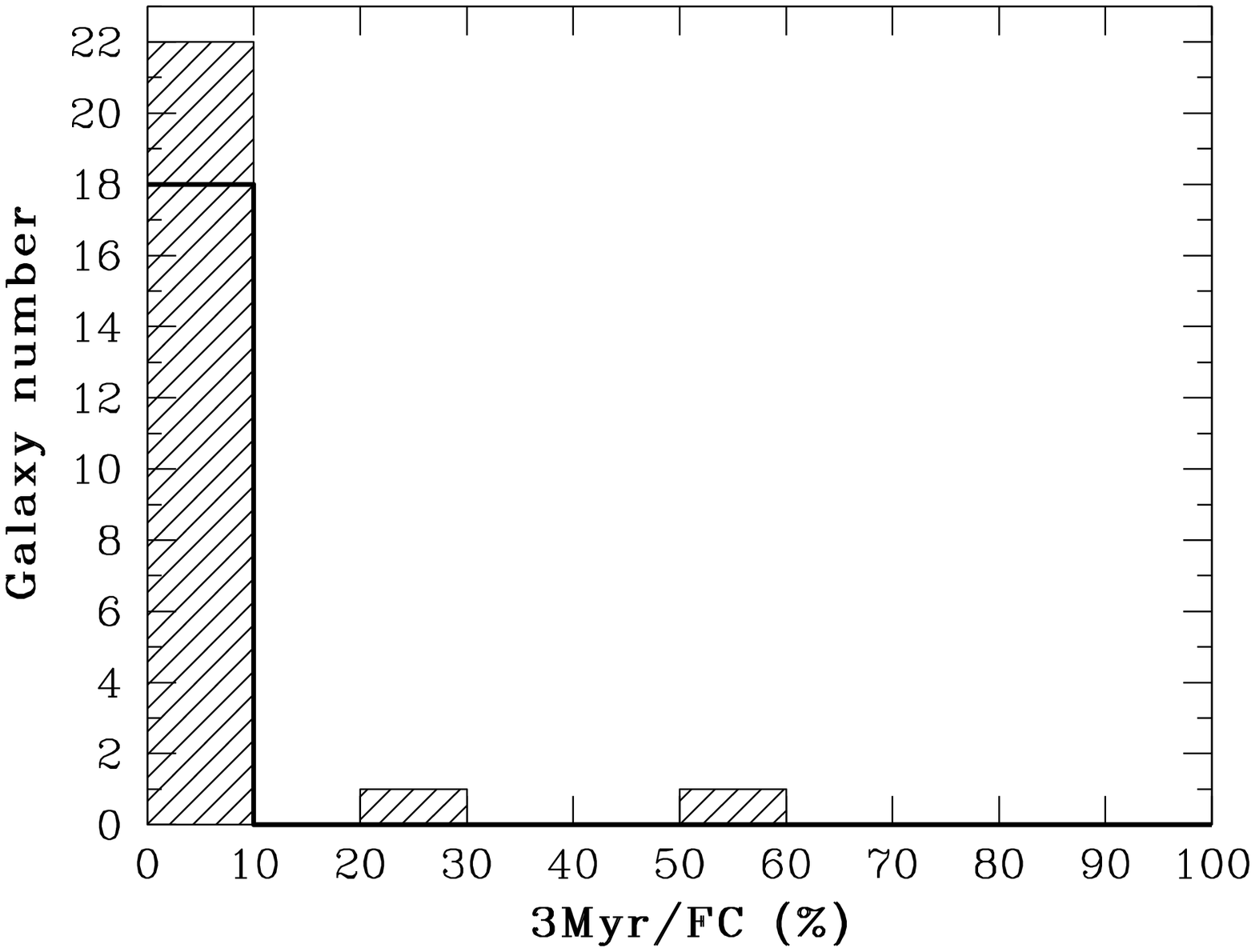}
\end{figure*}

Figure \ref{histo_radioxnaoativa_1.5kpc} shows comparative histograms for
the radio galaxies and control sample galaxies in the percentage
contributions of each age component to the flux at 4020\,\AA\ of the
nuclear spectra. In the lower half of Table \ref{res_sin_radio} we list
the mean contributions and corresponding standard deviations.

The histograms and Table \ref{res_sin_radio} show that the main
differences between the stellar populations of the radio galaxies and
control sample are in the relative contributions of the 10\,Gyr and
1\,Gyr components. 
%
In the radio galaxies, the contribution of the 10\,Gyr component is
systematically smaller, while the contribution of the 1\,Gyr component is
systematically larger than in the control sample.

The above results suggest that 1\,Gyr old star formation episodes have
been more frequent in radio galaxies than in non-active galaxies of the
same Hubble type, suggesting a relation between star-formation episodes
triggered 1\,Gyr ago and the presence of radio activity at the present
time. In addition, the correlation between the radio power and the
contribution of the 1\,Gyr component that we found in the previous
section suggests that the radio power is correlated with the mass of the
starburst.




%

\section{Summary and conclusions}

In this paper we have studied the nuclear and extranuclear stellar
populations of a complete sample of 24 radio galaxies and a control
sample of 18 non-active early-type galaxies, matched in Hubble type to
those of the radio sample. The main conclusions of this work can be
summarized as follows.

\smallskip\noindent
{\it i}\/) In most radio galaxies, the stellar population is dominated
by old (10\,Gyr) and intermediate age (1\,Gyr) components.

\smallskip\noindent
{\it ii}\/)  Blue continua due to an AGN and/or to recent star
formation episodes (100\,Myr old or younger) are found in only four of
the 24 radio galaxies. Two of them are BLRGs, for which the blue
continuum is probably dominated by the AGN light. The frequency of
clear signatures of recent star formation as the dominant source of blue
continuum in our complete sample is thus no more than 10-15\,\%. This
value is much smaller than the $\sim 40$\,\% frequency that we found in
our previous studies of Seyfert 2 galaxies (e.g., Storchi-Bergmann et
al.\ 2001, Raimann et al.\ 2003 and references therein).

\smallskip\noindent
{\it iii}\/) The four radio galaxies with a significant contribution from
young components and/or direct AGN light are all FR\,II/x. There seems,
therefore, to be a systematic difference between FR\,I and FR\,II stellar
populations, in the sense that FR\,II radio galaxies have larger
contributions from younger stellar populations. A larger sample is needed
to better quantify this difference.

\smallskip\noindent
{\it iv}\/) There is an inverse relation between the strength of the
emission line [O\,{\sc iii}] $\lambda$5007 and the mean age of the
stellar population, suggesting that younger galaxies have more active
nuclei.

\smallskip\noindent
{\it v}\/) There is also a relation between the radio power and the mean
age of the stellar population in the sense that younger galaxies host
more powerful radio sources.




\smallskip\noindent
{\it vi}) The main difference between the stellar populations of the
radio and control sample galaxies is that the former have, on average, a
larger contribution from the intermediate 1\,Gyr age component.
This excess contribution suggests a relation between
the present radio activity and a past episode of star formation which
occurred about 1\,Gyr ago.

\smallskip In order to explain the above results we speculate on an
evolutionary scenario for the radio galaxies, similar to the one we have
proposed for Seyfert galaxies (Storchi-Bergmann et al.\ 2001): a past
event which occurred more than $10^9$\,yrs ago (e.g., interaction
with a passing external galaxy, or merger) has triggered an episode of
star formation in the inner region. After an average delay of $10^9$\,yrs
the radio activity sets in. Our results also suggest that the more
massive the starburst, the stronger the subsequent radio emission.

\smallskip Comparing the present results for the radio galaxies with
those for Seyferts (Storchi-Bergmann et al.\ 2001), we conclude that
while for Seyfert\,2 galaxies the delay between the triggering of
star formation and the onset of activity would be, on average
10$^8$\,yr, for the radio galaxies this delay would be an order of
magnitude longer.

\section*{Acknowledgments}

DR and TSB acknowledge support from the Brazilian Institutions CNPq,
CAPES and FAPERGS.  RWH acknowledges support from grants awarded by the 
Australian Research Council. HQ acknowledges partial support from the 
FONDAP Centro de Astrofisica. We thank the referee for useful suggestions
which helped to improve the paper.

\label{lastpage}

\end{document}